\pdfoutput=1
\documentclass[aps,prd,amsmath,floatfix, twocolumn,
superscriptaddress,nofootinbib,showpacs,longbibliography]{revtex4-1}
\usepackage[T1]{fontenc}
\usepackage[utf8]{inputenc}
\usepackage{txfonts}
\usepackage{verbatim}
\usepackage{fontawesome5}
\usepackage[dvipsnames, usenames]{xcolor}
\definecolor{linkcolor}{rgb}{0.0,0.3,0.5}
\usepackage[hypertexnames=false, unicode, colorlinks=true, linkcolor=linkcolor,
citecolor=linkcolor, filecolor=linkcolor,urlcolor=linkcolor,
pdfusetitle]{hyperref}
\usepackage[all]{hypcap}
\usepackage{graphicx}
\usepackage{xspace}
\usepackage{amssymb}
\usepackage{microtype}
\usepackage{subfigure}
\usepackage{url}
\usepackage[english]{babel}
\usepackage{blindtext}
\usepackage[normalem]{ulem} 
\usepackage{bm} 
\usepackage{mathrsfs}
\usepackage{xspace} 

\usepackage{lipsum}

\DeclareMathAlphabet{\mathpzc}{OT1}{pzc}{m}{it}

\usepackage[nomessages]{fp}

\newcommand{\sk}[1]{}

\begin{document}
\title{Analysis of late-time tails in spin-aligned eccentric binary black hole mergers}

\newcommand{\KITP}{\affiliation{Kavli Institute for Theoretical Physics, University of California Santa Barbara, Kohn Hall, Lagoon Rd, Santa Barbara, CA 93106}} 
\author{Tousif Islam}
\email{tousifislam@ucsb.edu}
\KITP

\author{Guglielmo Faggioli}
\affiliation{Max Planck Institute for Gravitational Physics (Albert Einstein Institute), Am M¨uhlenberg 1, Potsdam, 14476, Germany}

\author{Gaurav Khanna} \affiliation{Department of Physics and Institute for AI \& Computational Research, University of Rhode Island, Kingston, RI 02881}
\affiliation{Department of Physics and Center for Scientific Computing \& Data Science Research, University of Massachusetts, Dartmouth, MA 02747}

\hypersetup{pdfauthor={Islam et al.}}
\date{\today}


\begin{abstract}
We present a comprehensive analysis of late-time tails in gravitational radiation from merging spin-aligned eccentric binary black holes, using high-accuracy point-particle black hole perturbation theory simulations. 
We simulate the late-time evolution of 15 binary black hole mergers with mass ratio $q = 1000$, dimensionless spins $\chi = [-0.9, -0.6, 0.0, 0.6, 0.9]$ and eccentricity at the last stable orbit $e_{\rm LSO} = [0.8, 0.9, 0.95]$. 
We track the tail amplitudes and exponents up to a retarded time coordinate $t = 9000M$ after merger for the six spin-weighted spherical harmonic modes $(2,1)$, $(2,2)$, $(3,2)$, $(3,3)$, $(4,3)$, and $(4,4)$ employing both frequentist and Bayesian approaches. We note that the tails are increasingly pronounced for binaries with high eccentricity $e_{\rm LSO}$ and large negative spin $\chi$.
We find that the overall late-time exponents closely approach their predicted asymptotic values ($p=-\ell-4$ for Weyl curvature scalar $\psi_{4,\ell m}$ where $\ell$ is the spin-weighted spherical harmonic index), while estimates restricted to the latest portion of the data exactly recover them.
We further verify numerically that modes with the same spherical index $\ell$ share identical tail exponents, while variations in $m$ do not affect the tail behavior. Additionally, for completeness, we provide a brief exploration of the intermediate behavior of the tails.
Our analysis framework is publicly available through the \texttt{gwtails} Python package~\footnote{\href{https://github.com/tousifislam/gwtails}{https://github.com/tousifislam/gwtails}}.
\href{https://github.com/tousifislam/gwtails}{\faGithub}
\end{abstract}

\maketitle
\section{Introduction}
\label{sec:intro}
The coalescence of a binary black hole (BBH) system can be divided into three distinct stages: \textit{inspiral}, \textit{merger}, and \textit{ringdown}. During the inspiral phase, the two black holes (BHs) orbit each other at relatively large separations, gradually losing energy and angular momentum through the emission of gravitational waves (GWs). 
At merger, the BHs plunge toward each other and their individual horizons coalesce to form a single, highly distorted remnant. In the subsequent ringdown phase, this remnant BH relaxes toward a stationary Kerr configuration by radiating away residual distortions in the form of quasi-normal mode (QNM) oscillations. Over the past decades, these three stages of BBH coalescence have been investigated in detail using both linear perturbative techniques (such as point-particle BH perturbation theory (ppBHPT)) and fully nonlinear numerical relativity (NR) simulations that track the dynamical evolution of the merging BHs~\cite{London:2018nxs,Berti:2014fga,Berti:2005ys,Berti:2005ys,Baibhav:2023clw,London:2018gaq,Redondo-Yuste:2023seq}.

Relatively less explored is the radiation that persists after QNMs have decayed. Perturbative analyses first revealed that this late-time stage of radiation is governed by a power-law decay, commonly referred to as the \textit{late-time tails} or \textit{Price tails}~\cite{Price:1971fb, Price:1972pw}. These tails arise due to the backscattering of GWs off the long-range curvature of the underlying spacetime. Since their discovery, late-time tails have been studied primarily using perturbative techniques in both Schwarzschild and Kerr spacetimes~\cite{Gundlach:1993tp, Gundlach:1993tn,Okuzumi:2008ej,Burko:1997tb,Barack:1998bw,Bernuzzi:2008rq,Burko:2013bra, Zenginoglu:2012us, Burko:2010zj, Burko:2004jn, Burko:2007ju, Krivan:1999wh, Poisson:2002jz, Burko:2002bt, Barack:1999ma,Racz:2011qu, Harms:2013ib,Zenginoglu:2009ey}. Because these signals become significant only at much later times after the merger and are several orders of magnitude weaker than the merger amplitudes, this regime has remained largely inaccessible to NR until recently. 
On the other hand, since the late-time tails are associated with the relaxation of the black hole formed at the end of the merger, they lie within the regime of validity of the ppBHPT framework. While the linear ppBHPT framework omits nonlinear corrections to the ringdown signal~\cite{Mitman:2022qdl,Cheung:2022rbm}, such contributions should remain subdominant for highly asymmetric binary mergers and may therefore be safely neglected.

Recently, interest in \textit{late-time tails} has been reignited by an intriguing result reported in Ref.~\cite{Albanesi:2023bgi,DeAmicis:2024not}, which observed that the late-time tail arises much earlier and with a more pronounced amplitude in the presence of orbital eccentricity. That study employed perturbative techniques within the Regge–Wheeler–Zerilli (RWZ) framework~\cite{PhysRev1081063,PhysRevLett24737,PhysRevD71104003,Nagar:2005ea} and was restricted to non-spinning BBH systems with eccentric orbits. In Ref.~\cite{Islam:2024vro}, we extended this analysis to both spin-aligned and non-spinning eccentric binaries using ppBHPT within the Teukolsky formalism~\cite{Sundararajan:2007jg,Sundararajan:2008zm,Sundararajan:2010sr,Zenginoglu:2011zz,WENO,Taracchini:2013wfa,Taracchini:2014zpa,Barausse:2011kb,Nagar:2006xv}. Our results reaffirmed the presence of enhanced late-time tails in both cases. Subsequently, these findings have been followed up in highly eccentric as well as head-on NR simulations of comparable-mass binaries, which confirmed the existence of pronounced late-time tails~\cite{DeAmicis:2024eoy,Ma:2024hzq}. 

From a theoretical perspective, late-time tails are particularly interesting because they provide a clean probe of the long-range structure of dynamical spacetimes and enable distinct predictions for the relaxation of compact objects within general relativity. At intermediate times, these tails also carry imprints of the binary dynamics, thereby offering potential avenues to infer properties of the progenitor BHs.

For the late-time tails, there exist clear theoretical predictions for their power-law decay. Specifically, for the GW strain $h(t)$, the tail exponents at future null infinity are expected to follow a scaling of $-(\ell + 2)$ for different spin-weighted spherical harmonic modes $(\ell, m)$. This implies that for the dominant quadrupolar mode, where the Weyl curvature scalar $\Psi_4 \propto \ddot{h}$, the exponent should be $-(\ell+4)$~\cite{Price:1971fb, Price:1972pw,Barack:1999st,Hod:1999ry}. At intermediate times, however, the observed tails, both in perturbative calculations and in NR simulations, exhibit exponents that deviate slightly from these asymptotic predictions. Such deviations can arise for two primary reasons. First, the perturbed BH may not have completely relaxed to its final stationary state. Second, additional intermediate tails, sourced by the excitation and mixing of QNMs with different decay rates, may dominate in this transitional regime. The existence of such intermediate tails has been investigated in detail in Refs.~\cite{Zenginoglu:2012us,Cardoso:2024jme,Okuzumi:2008ej}.

\begin{figure*}
\centering
\subfigure[]{
    \includegraphics[width=0.9\textwidth]{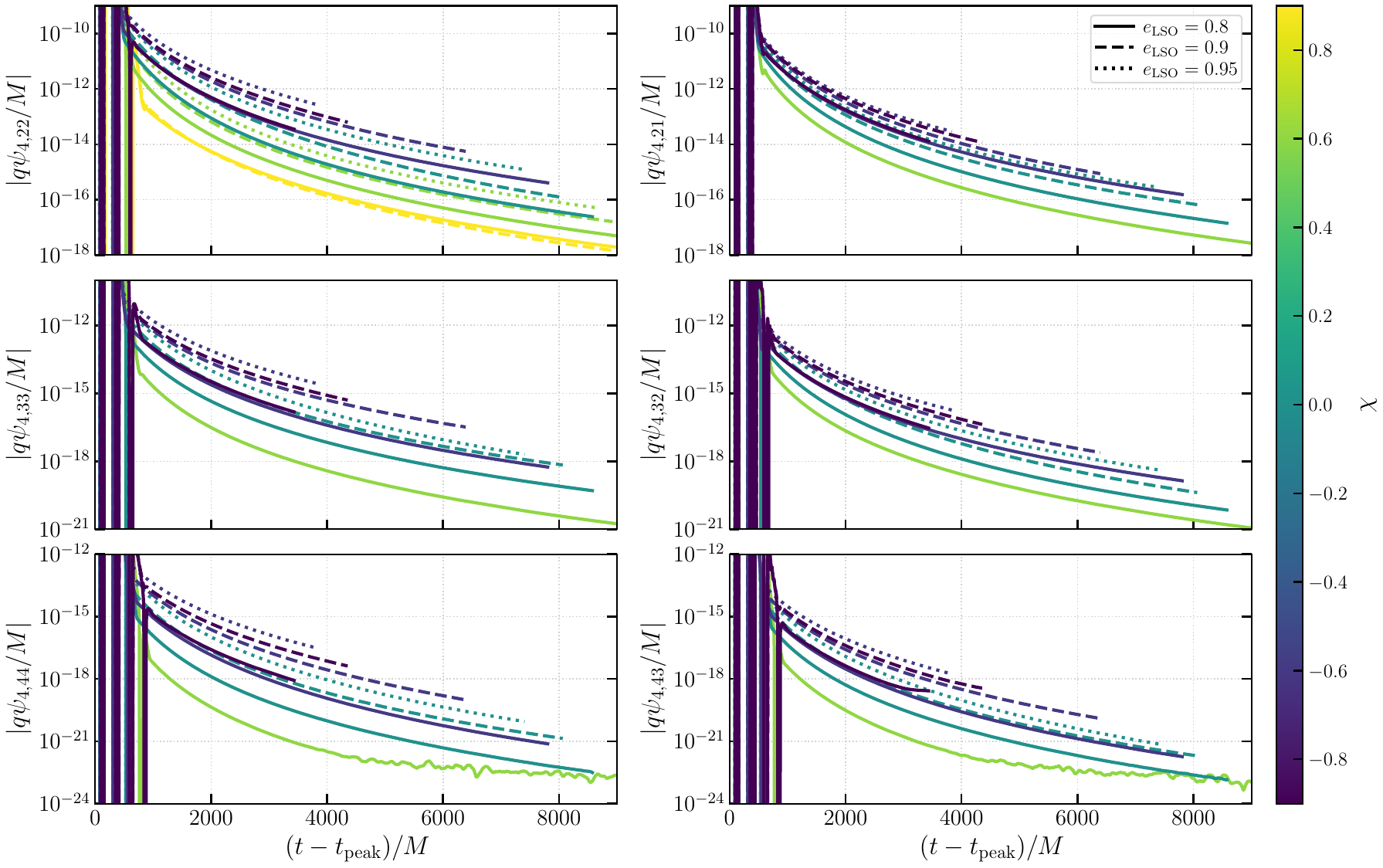}
    \label{fig:multi_mode_tails}
}
\vspace{0.3cm}
\subfigure[]{
    \includegraphics[width=0.9\textwidth]{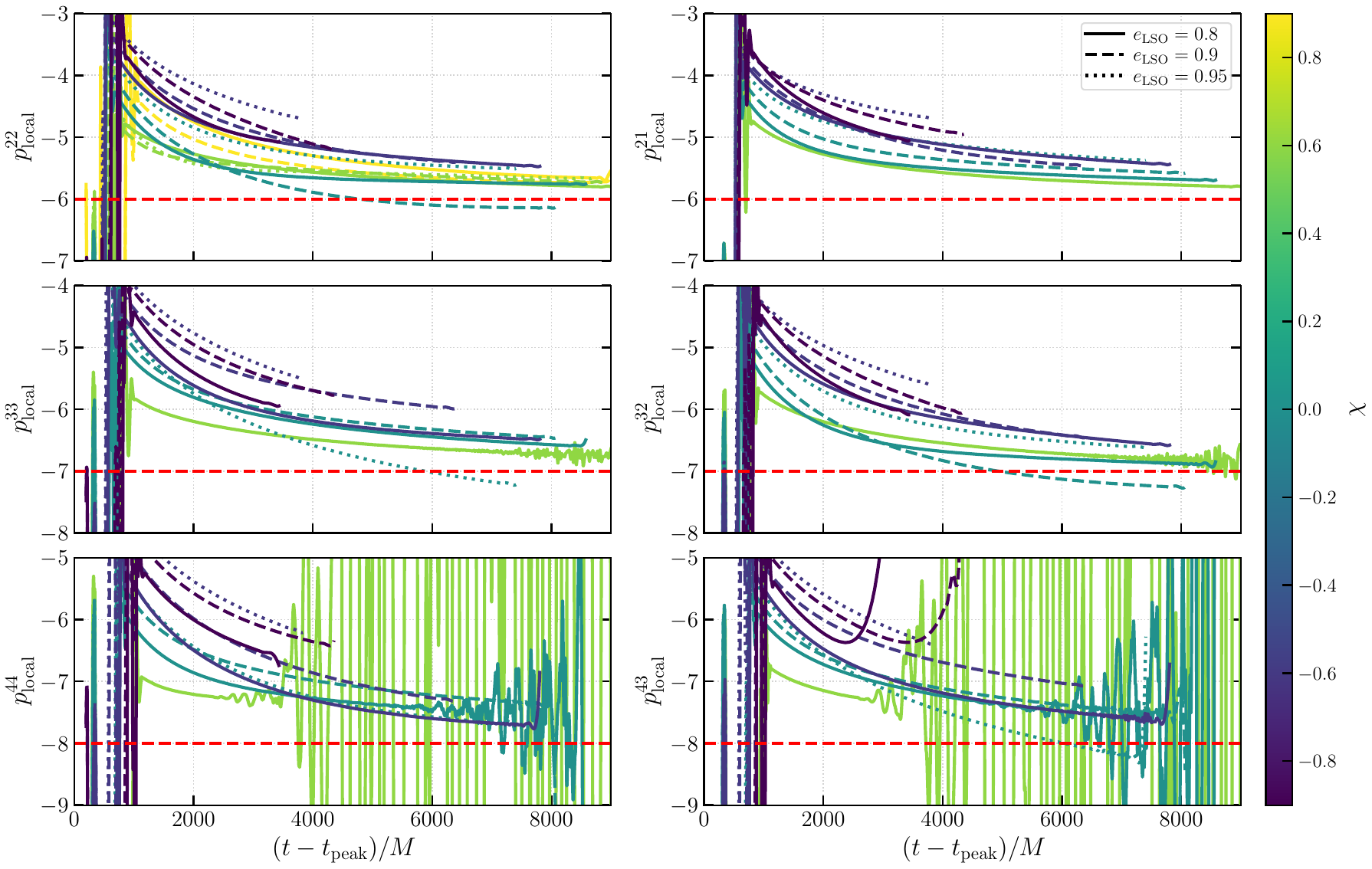}
    \label{fig:multi_mode_p_local}
}
\caption{We show the tail amplitudes (\textit{upper panel (a)}) and the local tail exponents (\textit{panel (b)}) for six representative modes, $[(2,1), (2,2), (3,2), (3,3), (4,3), (4,4)]$, corresponding to binaries with varying spins and eccentricities. The lines are color-coded according to the spin values. In panel (b), the expected asymptotic values are also shown as red horizontal dashed lines. We find that the tail exponents gradually approach their expected asymptotic values as the post-merger evolution progresses. More details are in Section~\ref{sec:result}.}
\label{fig:multi_mode_combined}
\end{figure*}

All previous studies, however, have primarily focused on the late-time tails in the quadrupolar modes. An exception is Ref.~\cite{DeAmicis:2024not}, which also presented the tail behavior for the $(\ell, m) = [(3,2), (4,4)]$ modes in the non-spinning case. In contrast, Refs.~\cite{DeAmicis:2024eoy,Ma:2024hzq} examined the tails only in the quadrupolar mode within NR simulations. One possible reason for the limited exploration of higher-order modes is that their amplitudes are expected to be smaller than those of the quadrupolar tails and may fall below the numerical noise floor in NR simulations. In Ref.~\cite{Islam:2024vro}, using the Teukolsky formalism, we simulated the late-time amplitudes for several higher-order modes but were able to evolve the system only up to approximately $1000M$ after merger. We found that, by this time, the post-merger amplitudes in the quadrupolar mode had already reached the level of numerical error, preventing us from probing the tail behavior further in that mode. For the higher-order modes, we observed that the numerical noise floor was reached even earlier, within a few hundred $M$ after the merger, making it impossible to identify any clear tail signatures.

Since then, several numerical improvements have been incorporated into the Teukolsky solver code~\cite{Sundararajan:2007jg,Sundararajan:2008zm,Sundararajan:2010sr,Zenginoglu:2011zz,WENO,Taracchini:2013wfa,Taracchini:2014zpa,Barausse:2011kb,Nagar:2006xv}, used in Ref.~\cite{Islam:2024vro}, including support for higher-precision floating-point arithmetic~\cite{Burko:2016sfi}. Furthermore, we chose to work directly with the Weyl curvature scalar $\Psi_4$, which is computed from the Teukolsky solver, rather than with the GW strain $h$, which requires a double time integration of $\Psi_4$. This approach eliminates potential numerical inaccuracies in our results. Because the tail amplitudes are extremely small, even minor numerical errors, typically negligible in other contexts, can hinder the accurate characterization of the tails, especially for the higher-order modes. 

With these new improvements, we now simulate the late-time evolution up to $9000M$ after merger for 15 eccentric, spin-aligned BBH systems with mass ratio $q = 1000$~\footnote{where $q = m_1/m_2$ with $m_1$ and $m_2$ being the mass of the larger and smaller black hole respectively}, dimensionless spin parameters as high as $0.9$ and eccentricities at the last stable orbit (LSO) up to $0.95$. These simulations exhibit clear signatures of late-time tails across several spin-weighted spherical harmonic modes. For simplicity, we focus on six spin-weighted spherical harmonic modes, $(2,1)$, $(2,2)$, $(3,2)$, $(3,3)$, $(4,3)$, and $(4,4)$, two for each spherical harmonic index $\ell$. We analyze the behavior of these tails using both frequentist and Bayesian approaches. First, we observe that the tails are most pronounced for binaries with high eccentricity $e_{\rm LSO}$ and large negative spin $\chi$.
Our results also show that the late-time decay exponents closely approach the theoretically predicted asymptotic values ($p = -\ell - 4$ for $\psi_{4,\ell m}$), while estimates restricted to the latest portion of the signal recover them exactly. We also confirm numerically that modes sharing the same spherical index $\ell$ exhibit identical tail exponents, indicating that variations in $m$ do not affect the decay behavior. The entire analysis is performed using the recently developed \texttt{gwtails} package~\cite{Islam:2024vro}, which is publicly available at \href{https://github.com/tousifislam/gwtails}{\texttt{https://github.com/tousifislam/gwtails}}.

Ref.~\cite{Becker:2025inprep}, completed contemporaneously with this work, investigates how the anomaly parameter—required, together with the eccentricity parameter, for a complete description of eccentric motion—affects the tail behavior of the quadrupolar mode, using the same Teukolsky code and the \texttt{gwtails} package (with least-squares fits).
They find that variations in the anomaly parameter, in combination with variations in the eccentricity, significantly modify the inferred tail parameters in a manner similar to how these variations modify the spectrum of quasi-normal modes.  Taken together, their results indicate that the precise nature of a source’s late-time dynamics, which depend on both eccentricity and anomaly parameter, have a large and important influence on tails.
In our analysis, however, we fix the anomaly value. Taken together, this paper and Ref.~\cite{Becker:2025inprep} provide a new perspective on how tails depend on binary parameters and on how tail parameters can be extracted in a robust and reliable manner.

\section{Methodology}
\label{sec:methodology}
We now provide the outline of our methodology. We first describe how the waveforms emitted by the test-particle are computed at future null infinity by sourcing the time-domain Teukolsky code~\cite{Sundararajan:2007jg,Sundararajan:2008zm,Sundararajan:2010sr,Zenginoglu:2011zz,WENO,Burko:2016sfi} with the particle's trajectories. Then, we describe how we model the late-time tails, and finally, we provide details on the fitting procedure employed in the modeling of the late-time tails. Throughout this paper, we adopt geometric units $G = c = 1$.

\subsection{Trajectories and waveforms computation}
We employ the ppBHPT framework to simulate mergers of two BHs with masses $m_1$ and $m_2$, where $m_2 \ll m_1$, or equivalently, with a mass ratio $q = m_1/m_2 \gg 1$. In the ppBHPT approach, the smaller BH is modeled as a structureless point particle moving in the spacetime of a larger Kerr BH with dimensionless spin parameter $\chi = J/M^2$, where $J$ denotes the intrinsic angular momentum of the Kerr BH and $M = m_1 + m_2$ is the total mass of the system. The particle acts as a perturbation to the Kerr metric, that we express in Boyer–Lindquist coordinates $\{\tilde{t}, \tilde{r}, \theta, \varphi \}$. Furthermore, we consider scaled dimensionless variables $t=\tilde{t}/M$ and $r=\tilde{r}/M$. This formulation allows us to compute the GWs emitted by the particle by solving the Teukolsky equation~\cite{Teukolsky:1973} in the time domain, using the code developed in Refs.~\cite{Sundararajan:2007jg,Sundararajan:2008zm,Sundararajan:2010sr,Zenginoglu:2011zz,WENO,Burko:2016sfi}.

The first step in computing the GWs is to determine the trajectory of the point particle. To this end, we describe the dynamics of the particle orbiting the Kerr BH using the effective-one-body (EOB) formalism~\cite{Buonanno:1998gg,Buonanno:2000ef}. Within the EOB framework, at leading order in $1/q$, a BBH system described above can be mapped onto an effective system consisting of a point particle of mass $\mu$ orbiting a Kerr BH of mass $M$ and dimensionless spin $\chi$. Here, $\mu = m_1 m_2 / M$ denotes the reduced mass. For convenience, we set the mass of the Kerr BH to $M = 1$, unless stated otherwise, and fix $\mu = 10^{-3} M$, that corresponds to a BBH with mass ratio $q=1000$.

The particle’s trajectory is obtained by evolving the Hamiltonian system described in Eqs.~(5a)–(5d) of Ref.~\cite{Faggioli:2024ugn}. This system incorporates dissipative effects due to GW emission through a radiation-reaction (RR) force that includes resummed post-Newtonian (PN) eccentric corrections~\cite{Khalil:2021txt,Faggioli:2024ugn,Gamboa:2024imd}. We restrict our analysis to bound, eccentric, spin-aligned BBHs, which correspond to bound eccentric motion of a test particle confined to the equatorial plane of the Kerr BH. The eccentric motion is parametrized using the Keplerian parametrization in terms of the eccentricity $e$, semilatus rectum $p$, and relativistic anomaly $\xi$~\cite{Darwin:1959,Darwin:1961},  
\begin{align} \label{eq.: e, p, xi def}
	e &= \frac{r_a - r_p}{r_a + r_p}, \quad  
	p = \frac{2r_a r_p}{r_a + r_p}, \quad  
	\cos \xi = \frac{p - r}{e r} \, ,
\end{align}
where $r_a$ and $r_p$ denote the radial coordinates of the apocenter and pericenter, respectively. We initialize the system close to the transition to plunge, typically about two orbits before the LSO, and continue the evolution well beyond the coalescence to adequately capture the late-time tail behavior. During the plunge phase, following Ref.~\cite{Taracchini:2014zpa}, the RR force is smoothly switched off once the test particle reaches the inner light ring of the Kerr BH. Further details can be found in Sec.~II of Ref.~\cite{Taracchini:2014zpa}.  

We refer the values of the eccentricity and relativistic anomaly at the LSO, which corresponds to the last configuration for which these quantities are defined during the evolution. We generate a set of 15 trajectories with spin $\chi = [-0.9, -0.6, 0.0, 0.6, 0.9]$ and eccentricities $e_{\rm LSO} = [0.8, 0.9, 0.95]$, fixing the relativistic anomaly at the LSO to $\xi_{\rm LSO} = \pi$.

Once the trajectory of the test-particle is generated, we solve the inhomogeneous Teukolsky equation in the time domain, supplying the computed trajectory as a source term~\cite{Sundararajan:2007jg,Sundararajan:2008zm,Sundararajan:2010sr,Zenginoglu:2011zz,WENO,Burko:2016sfi}. The numerical implementation proceeds in several steps: 
(i) the Teukolsky equation is reformulated using compactified hyperboloidal coordinates, enabling waveform extraction directly at future null infinity and addressing the outer boundary problem of the computational grid;  
(ii) axisymmetry of the Kerr background is exploited to reduce the system to a set of $(2+1)$-dimensional partial differential equations (PDEs);  
(iii) these equations are recast into a first-order hyperbolic system; and  
(iv) the resulting system is evolved using a second-order Lax-Wendroff finite-difference scheme with explicit time integration. The Weyl scalar $\Psi_4$ as a function of the retarded time $t$ is then extracted at future null infinity to obtain the radiative part of the gravitational field.

The Weyl scalar $\Psi_4$ can then be integrated twice to find the two polarization states $h_{+}$ and $h_{\times}$ of the transverse-traceless metric perturbations,
\begin{equation}
\Psi_4 = \frac{1}{2}\left(\frac{\,\partial^2h_{+}}{\,\partial t^2}-i\frac{\,\partial^2h_{\times}}{\,\partial t^2}\right)\, .
\end{equation}
The complex GW strain
\begin{align}
h_{+}((t, r, \theta, \varphi; \boldsymbol{\lambda}) &- {\mathrm i} h_{\times}((t, r, \theta, \varphi; \boldsymbol{\lambda}) \nonumber \\
& =  \sum_{\ell=2}^{\infty} \sum_{m=-\ell}^{\ell} h_{\ell m}(t, r, \theta, \varphi; \boldsymbol{\lambda}) {}_{-2}Y_{\ell m}(\theta, \varphi) \, ,
\label{eq:strain}
\end{align}
can be formed from the two polarization states,
which is subsequently decomposed into spin-weighted spherical harmonic modes of weight $s = -2$.
Unlike our previous work in Ref.~\cite{Islam:2024vro}, we choose to work with the the Weyl scalar $\Psi_4$ and decompose it into spin-weighted spherical harmonic modes, defined as
\begin{equation}
\Psi_4(t, r, \theta, \varphi; \boldsymbol{\lambda}) = \sum_{\ell=2}^{\infty} \sum_{m=-\ell}^{\ell} \psi_{4,\ell m}(t, r; \boldsymbol{\lambda})\; {}_{-2}Y_{\ell m}(\theta, \varphi).
\end{equation}
Here $\boldsymbol{\lambda}$ is the set of intrinsic parameters (such as the masses and spins of the binary) describing the binary, and ($\theta$,$\varphi$) are angles describing the orientation of the binary with respect to the observer. This decomposition enables us study the late-time tail behavior mode by mode.
Each complex spin-weighted spherical harmonic mode $\psi_{4, \ell m}(t)$ is further decomposed into a real amplitude $A_{\ell m}(t)$ and a phase $\phi_{\ell m}(t)$, defined as
\begin{equation}
\psi_{4,\ell m}(t) = A_{\ell m}(t)\, e^{i \phi_{\ell m}(t)} \, .
\label{eq:amp_phase}
\end{equation}
The time coordinate is chosen such that $t = 0$ corresponds to the instant of the last peak of the amplitude of the $(2,2)$ mode.

\subsection{Tail model}
The post-merger radiation can be described as a combination of QNM oscillations and late-time tail contributions. The QNMs correspond to exponentially damped sinusoids whose frequencies and damping times are determined by the mass and spin of the remnant BH~\cite{Teukolsky:1973,Detweiler:1980g,Leaver1985,Dolan:2009nk}. For each individual spherical harmonic mode, the total post-merger signal can be expressed as
\begin{equation}
    \psi^{\mathrm{post}}_{4, \ell m}(t)
    = \psi^{\mathrm{QNM}}_{4, \ell m}(t)
    + \psi^{\mathrm{tail}}_{4, \ell m}(t).
\end{equation}
The QNM component represents the superposition of all damped oscillatory contributions associated with the $(\mathfrak{l}, m, n)$ spin-weighted spheroidal harmonic modes~\cite{William:1973,Teukolsky:1973,Teukolsky1972},
\begin{align}
    \psi^{\mathrm{QNM}}_{4, \ell m}(t)
    &= \sum_{\mathfrak{l}=2}^{\infty}
    \sum_{n=0}^{\infty}
    \mathcal{A}_{\mathfrak{l} m n}\,
    e^{ -t / \tau_{\mathfrak{l} m n}
    - i \omega_{\mathfrak{l} m n} t
    - i \phi_{\mathfrak{l} m n} }
    \nonumber \\
    &\quad
    + \sum_{\mathfrak{l}=2}^{\infty}
    \sum_{n=0}^{\infty}
    \mathcal{A}'_{\mathfrak{l} m n}\,
    e^{ -t / \tau'_{\mathfrak{l} m n}
    - i \omega'_{\mathfrak{l} m n} t
    - i \phi'_{\mathfrak{l} m n} } .
\end{align}
Here, $\omega_{\mathfrak{l} m n}$ and $\tau_{\mathfrak{l} m n}$ denote the characteristic oscillation frequency and damping time of the $(\mathfrak{l}, m, n)$ QNM, while $\mathcal{A}_{\mathfrak{l} m n}$ and $\phi_{\mathfrak{l} m n}$ are its amplitude and phase, respectively. The primed quantities correspond to the so-called ``mirror modes'', and $n$ labels the overtone number. The fundamental mode ($n=0$) typically dominates the early ringdown signal~\cite{Berti:2005ys,Lim:2019xrb,Li:2021wgz}. 

\begin{figure*}[tb]
\includegraphics[width=\textwidth]{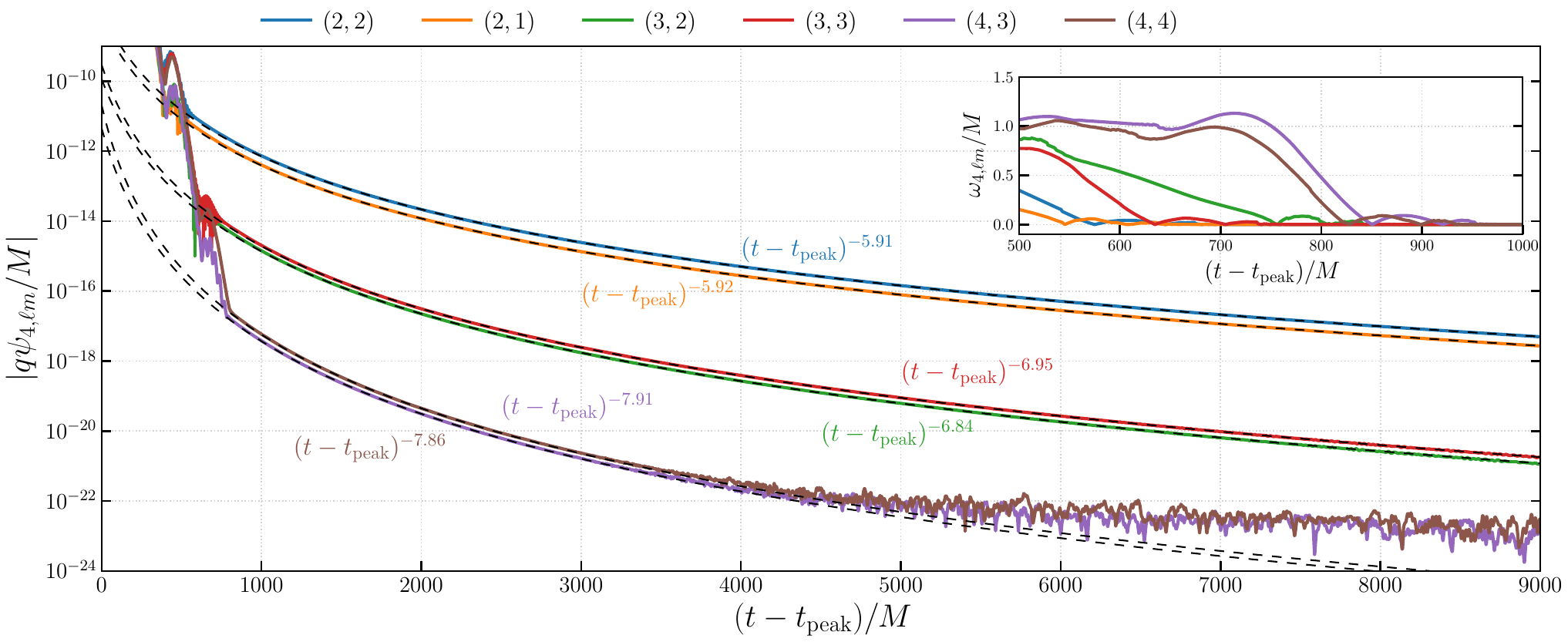}
\caption{We show the post-merger amplitudes and frequencies (inset) are shown for six representative modes, $[(2,1), (2,2), (3,2), (3,3), (4,3), (4,4)]$, corresponding to a binary characterized by $[\chi, e_{\rm LSO}] = [0.6, 0.8]$. For reference, we also indicate the best-fit exponents obtained through a Bayesian analysis performed with the \texttt{gwtails} package, which employs the \texttt{emcee} sampler. We find that the best-fit tail exponents are close to their expected asymptotic values. We find that the best-fit tail exponents are close to their expected asymptotic values. Furthermore, the post-merger frequencies $\omega_{\ell m}$ approach zero at different times for different spin-weighted spherical harmonic modes, indicating that the onset of the tail phase occurs at different times for each mode. More details are in Section~\ref{sec:result}. }
\label{fig:example_hms_tail}
\end{figure*}

\begin{figure*}
\centering
\subfigure[]{
    \includegraphics[width=0.45\textwidth]{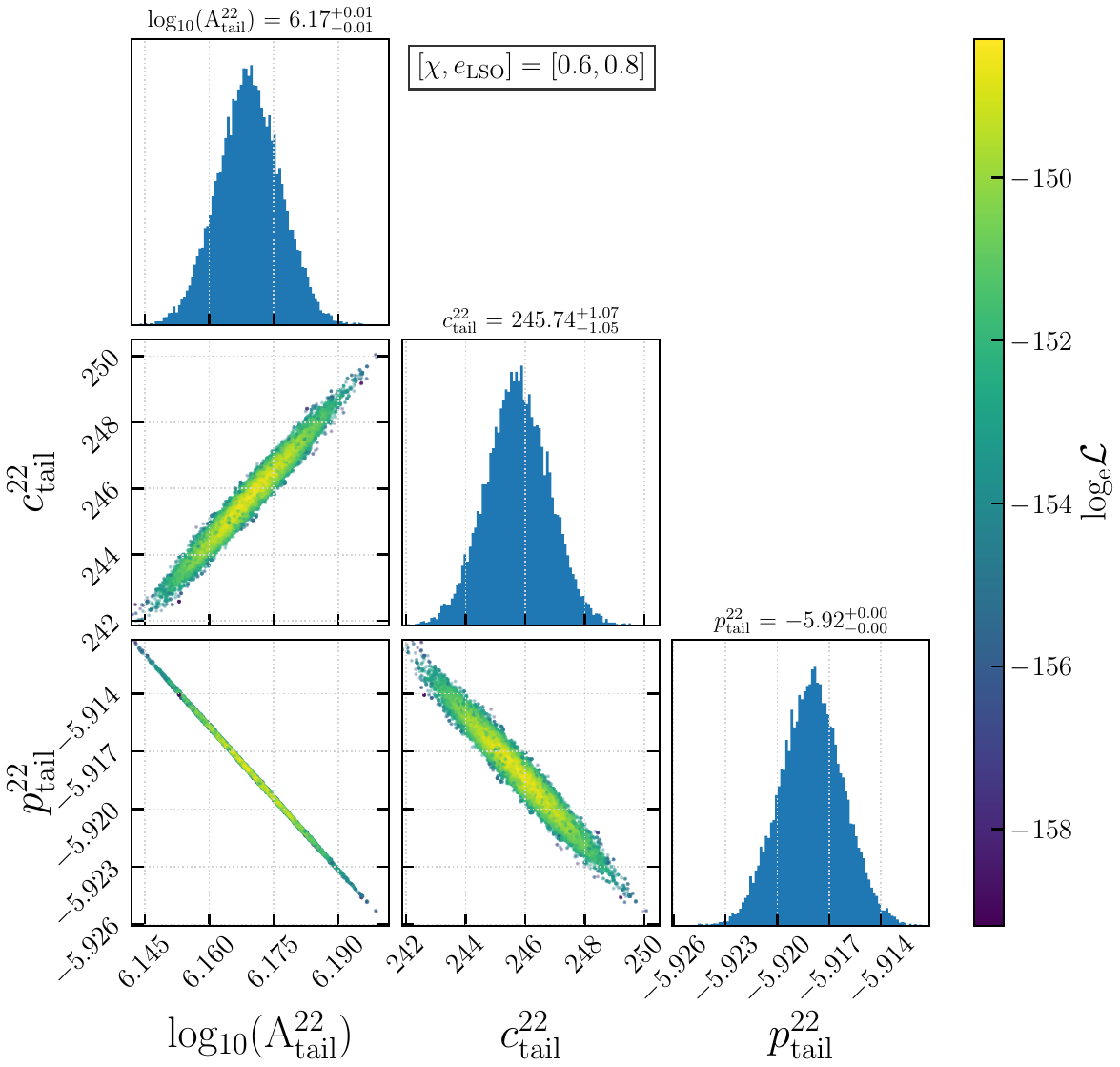}
    \label{fig:example_22_tail_mcmc}
}
\subfigure[]{
    \includegraphics[width=0.45\textwidth]{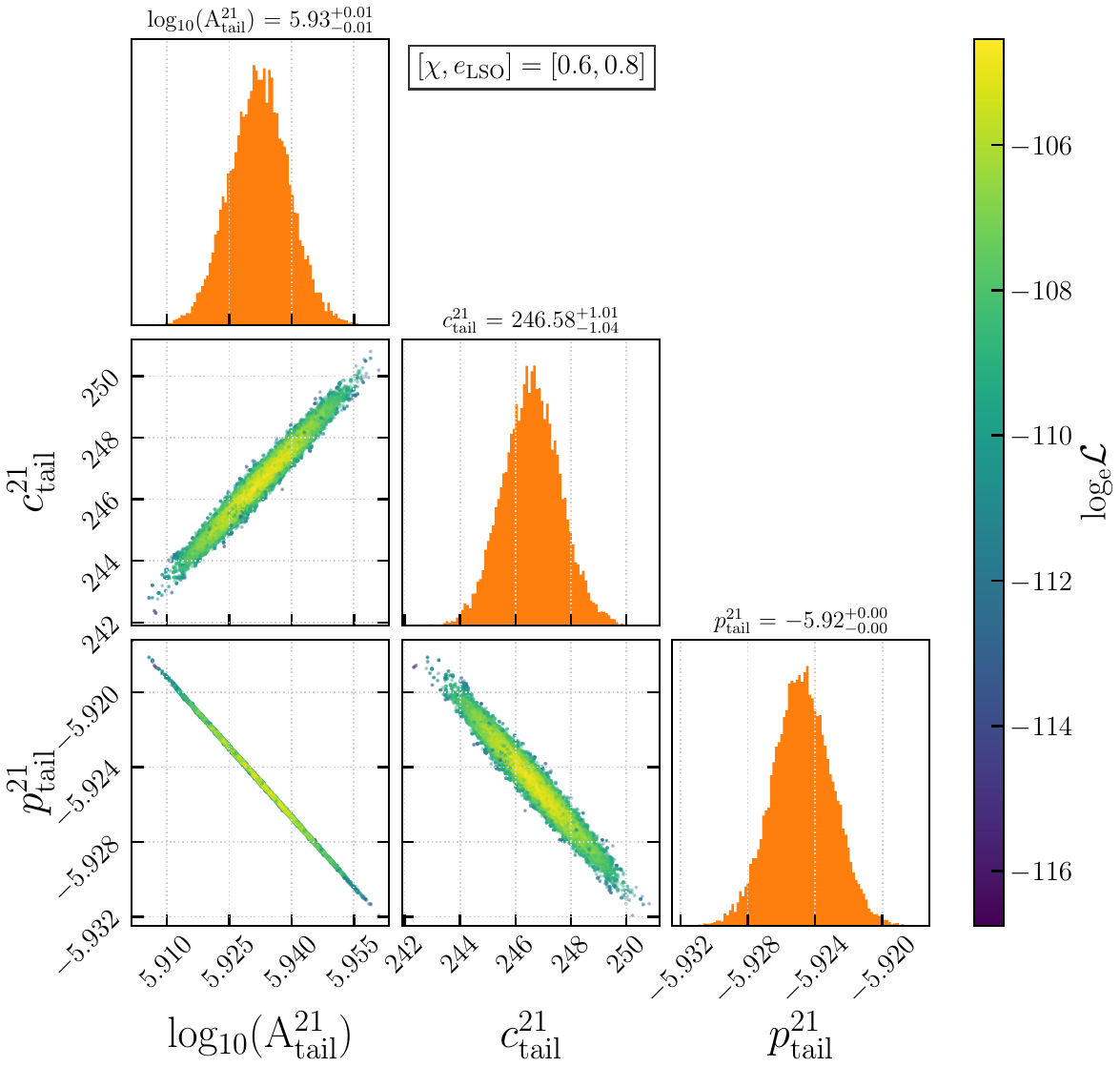}
    \label{fig:example_21_tail_mcmc}
}
\subfigure[]{
    \includegraphics[width=0.45\textwidth]{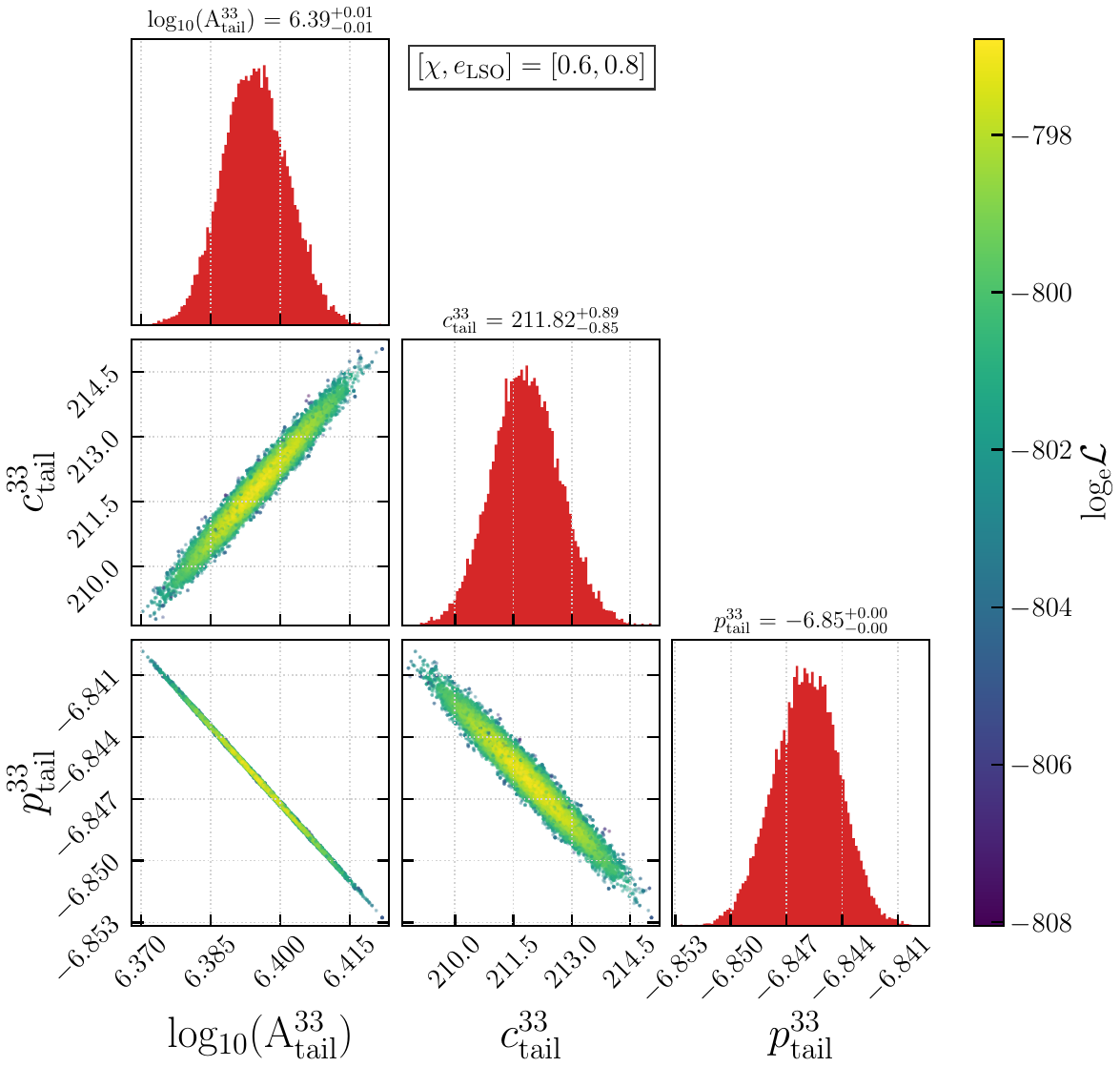}
    \label{fig:example_33_tail_mcmc}
}
\subfigure[]{
    \includegraphics[width=0.45\textwidth]{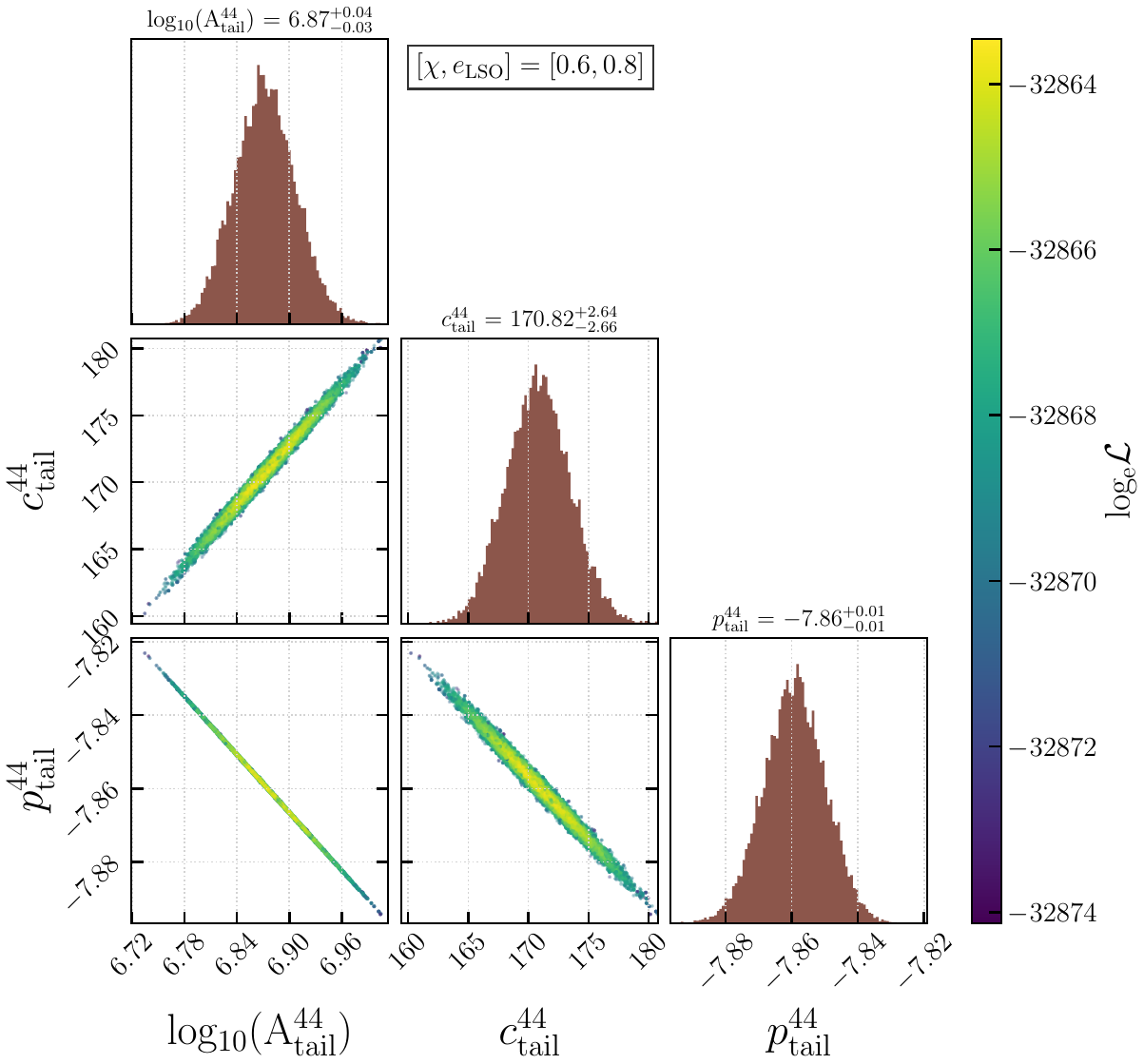}
    \label{fig:example_44_tail_mcmc}
}
\caption{We present the corner plot for the tail parameters $A_{\mathrm{tail}}^{\ell m}$, $p_{\mathrm{tail}}^{\ell m}$, and $c_{\mathrm{tail}}^{\ell m}$ obtained from Bayesian fits for four representative modes: $(2,1)$, $(2,2)$, $(3,3)$, and $(4,4)$. The results correspond to a binary characterized by $[\chi, e_{\rm LSO}, ] = [0.6, 0.8]$. The Bayesian analysis is performed using the \texttt{gwtails} package, which employs the \texttt{emcee} sampler for MCMC sampling. The analysis is performed over the time window $t = [1200, 8000]M$ for the $(2,1)$, $(2,2)$, and $(3,3)$ modes, while for the $(4,4)$ mode we restrict the fit to $t = [1200, 4000]M$ due to enhanced numerical noise at later times. The scatter points in each corner plot are color-coded according to their corresponding log-likelihood values. More details are in Section~\ref{sec:example_bbh}.}
\label{fig:example_tail_mcmc}
\end{figure*}

The $\psi^{\mathrm{tail}}_{4, \ell m}$ term captures the contribution from the late-time tails, which arise from the backscattering of GWs off the spacetime curvature at large radii. These tails originate from the branch cut of the Green’s function along the zero-frequency axis~\cite{Casals:2012tb,Ching:1995tj}. At sufficiently late times, their contribution to the waveform can be modeled as
\begin{equation}
    \psi^{\mathrm{tail}}_{4,\ell m}(t)
    = A_{\mathrm{tail}}^{\ell m}
    (t + c_{\mathrm{tail}}^{\ell m})^{p_{\mathrm{tail}}^{\ell m}}
    e^{i \phi_{\mathrm{tail}}^{\ell m}},
    \label{eq:htail}
\end{equation}
where the decay exponent is theoretically predicted to be
$p_{\mathrm{tail}}^{\ell m} = -(\ell + 4)$~\cite{Barack:1999st,Hod:1999ry}.

In contrast to Ref.~\cite{Islam:2024vro}, where the post-merger radiation have been evolved only up to $\sim 1000M$ after the merger, we can now simulate the waveforms out to $\sim 9000M$. This extended evolution enables us to access the regime that is expected to be fully dominated by the late-time tail. In particular, the early portion of the radiation (up to $\sim 100M$--$200M$) is primarily QNM-dominated, while the interval between $\sim 100M$ and $\sim 500M$ corresponds to an intermediate phase where the leading-order QNMs and the tail contribution have comparable amplitudes, leading to noticeable interference~\cite{Albanesi:2023bgi,DeAmicis:2024not,DeAmicis:2024eoy,Ma:2024hzq,Islam:2024vro}. We therefore focus exclusively on times $t \geq 1200M$, which lie well beyond both the QNM-dominated and the intermediate regimes where QNMs and tails can overlap. In this tail-dominated region, the amplitude of $\psi^{\mathrm{tail}}_{4,\ell m}(t)$ is modeled by
\begin{equation}
|\psi^{\mathrm{tail}}_{4,\ell m}(t)|
= A_{\mathrm{tail}}^{\ell m}\,(t + c_{\mathrm{tail}}^{\ell m})^{p_{\mathrm{tail}}^{\ell m}},
\label{eq:tailmodel}
\end{equation}
where $A_{\mathrm{tail}}^{\ell m}$ represents the overall amplitude scale, $p_{\mathrm{tail}}^{\ell m}$ is the decay exponent governing the power-law falloff, and $c_{\mathrm{tail}}^{\ell m}$ denotes an effective time offset corresponding to the onset of the tail-dominated regime.

\begin{figure*}
    \centering
    \subfigure[]{
        \includegraphics[width=0.51\textwidth]{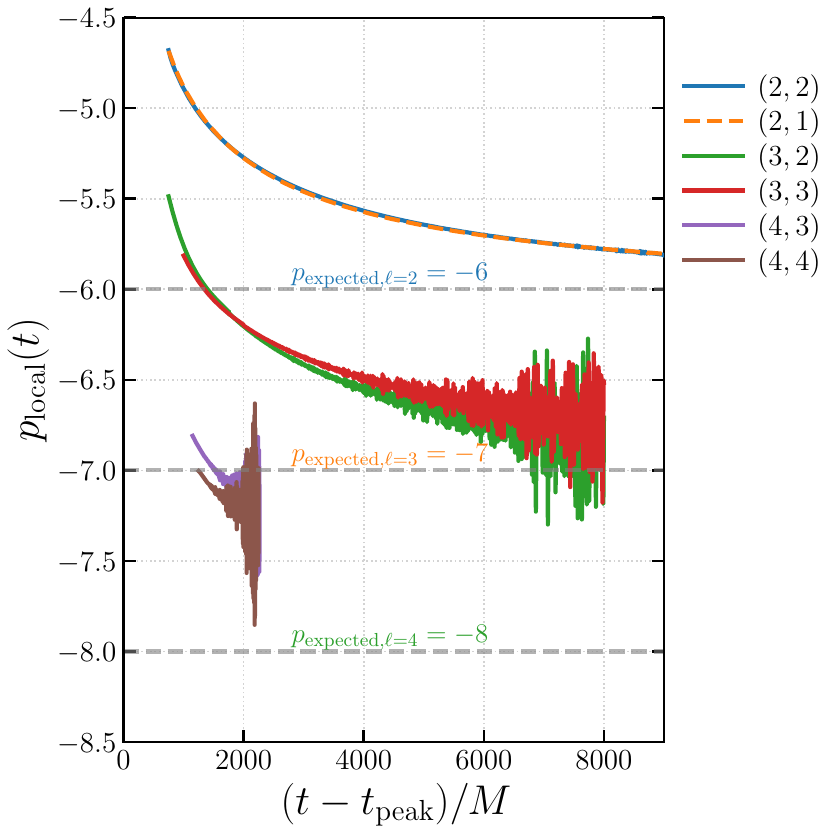}
        \label{fig:example_pt_tail}
    }
    \subfigure[]{
        \includegraphics[width=0.43\textwidth]{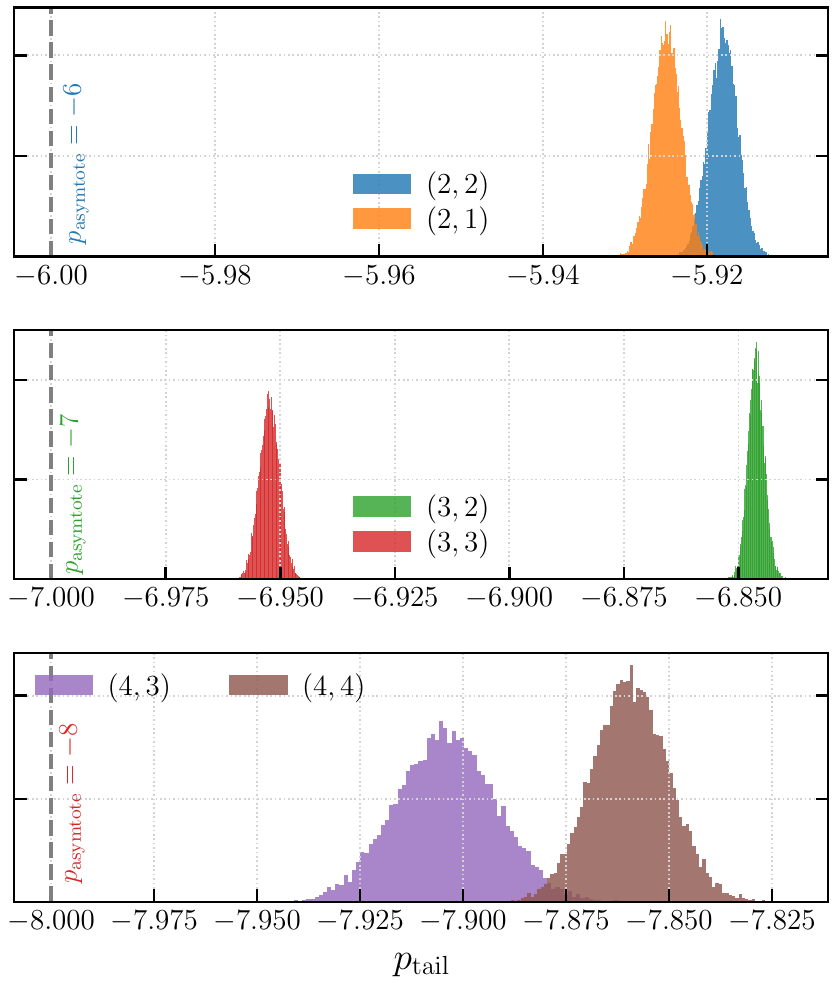}
        \label{fig:example_ptail_hists}
    }
    \caption{We show the local tail exponents $p(t)$ (\textbf{Left}) and histogram of the estimated tail exponent $p_{\rm tail}^{\ell m}$ using Bayesian sampler (\textbf{Right}) for six representative modes, $[(2,1),(2,2),(3,2),(3,3),(4,3),(4,4)]$, for a binary characterized by $[\chi, e_{\rm LSO}]=[0.6,0.8]$. The analysis is performed over the time window $t = [1200, 8000]M$ for the $(2,1)$, $(2,2)$, and $(3,3)$ modes, while for the $(4,4)$ mode we restrict the fit to $t = [1200, 4000]M$ due to enhanced numerical noise at later times. We find that not only are the tail exponents close to their expected asymptotic values, but modes with the same $\ell$ also yield sufficiently similar values. 
    For $\ell = 3$, we observe a slight discrepancy between the $p_{\mathrm{tail}}^{\ell m}$ values inferred from the $(3,3)$ and $(3,2)$ modes. We attribute this difference to enhanced numerical noise present in the late-time portion of the waveform, which is also visible in the evolution of $p_{\rm local}(t)$.
    More details are in Section~\ref{sec:example_bbh}. 
    }
    \label{fig:example_ptail_combined}
\end{figure*}

\subsection{Fitting framework}
We fit the late-time tails using the analytical model defined in Eq.~(\ref{eq:tailmodel}), implemented via the \texttt{gwtails}~\cite{gwtails} package. The details of our original computational framework are provided in Ref.~\cite{Islam:2024vro}. First, we perform frequentist fits using the \texttt{scipy.curve\_fit} routine~\cite{scipy}, which implements nonlinear least-squares (LSQ) optimization. This approach provides point estimates of the model parameters along with their covariance, offering a fast and computationally efficient method for initial parameter inference.

In addition, we incorporate Bayesian parameter estimation to quantify the full posterior distributions of the model parameters, accounting for potential correlations. For this purpose, we employ the \texttt{emcee} sampler~\cite{Foreman_Mackey_2013}, which implements the affine-invariant ensemble Markov chain Monte Carlo (MCMC) algorithm~\cite{2010CAMCS65G}. The Bayesian approach complements the frequentist analysis by providing robust uncertainty quantification and a more complete statistical characterization of the inferred parameters of the considered tail model in Eq.~\eqref{eq:tailmodel}.

For the Bayesian analysis, we assume a Gaussian likelihood function, since the Bayesian maximum a posteriori (MAP) estimate under a flat prior and a Gaussian likelihood with identical variance for all data points reduces to the frequentist least-squares estimate~\cite{Bishop2006,Pascal2012}. We adopt a conservative estimate of the numerical uncertainty in our simulations at the level of $2.5\%$, consistent with earlier error analyses (see, e.g., Fig.~6 of Ref.~\cite{Islam:2022laz} and Fig.~5 of Ref.~\cite{Rink:2024swg}) and set the variance in the Gaussian likelihood accordingly. 
Within \texttt{emcee}, we use \texttt{nwalkers = 50} and evolve each MCMC chain for $5000$ steps, discarding the initial $2500$ samples as the ``burn-in'' phase during which the walkers explore the parameter space before converging to the stationary distribution. Although a smaller burn-in would suffice, we conservatively remove half of the initial samples to ensure that transient effects do not influence the posterior distributions.

In our analysis, we adopt wide uniform priors~\footnote{While it is possible to explore other choices of likelihoods and priors, we have not performed such experiments in this work. A dedicated study could, in principle, be conducted in the future to assess the robustness of our results under varying likelihood and prior choices.} for each of the tail parameters $A_{\mathrm{tail}}^{\ell m}$, $c_{\mathrm{tail}}^{\ell m}$, and $p_{\mathrm{tail}}^{\ell m}$. Specifically, we set the prior ranges as $\log_{10}A_{\mathrm{tail}}^{\ell m} \in [0, 20]$, $c_{\mathrm{tail}}^{\ell m} \in [0, 2500]M$, and $p_{\mathrm{tail}}^{\ell m} \in [-15, -1]$. To initialize the MCMC sampling, we place the walkers around some fiducial parameter values. The sampler then explores the full parameter space to infer the posterior distributions and identify the best-fit parameters. As a consistency check, we also initialize the sampler near the best-fit values obtained from the LSQ fits and find that the resulting posteriors remain consistent with those obtained using the fiducial initialization.

We have now integrated our Bayesian fitting interface into the \texttt{gwtails} package, making it publicly available for use. As discussed earlier, in this work we focus explicitly on the late-time regime in order to clearly isolate the tail-dominated behavior. Unless stated otherwise, our default fitting window corresponds to $t \in [1200M, 8000M]$ after the merger.

\section{Late-time tails}
\label{sec:result}
We now present a detailed analysis of the late-time tails in six different spin-weighted spherical harmonic modes. Specifically, we select modes with different $m=\{l,l-1\}$ values for each $\ell = \{2, 3, 4\}$. While our simulations provide additional modes beyond those studied here, we focus on this subset because the remaining modes are subdominant and exhibit relatively larger numerical errors, making them less suitable for probing the late-time tail behavior reliably.

Before studying the post-merger data in more detail, we first inspect the amplitudes of these modes visually in Figure~\ref{fig:multi_mode_tails}. We find that, for certain modes, some simulations (specially with high eccentricities and high positive spins) exhibit unphysical behavior or significant numerical noise, and we discard such cases from our analysis. 
Furthermore, to mitigate numerical noise as much as possible before visual inspection, we apply a Savitzky–Golay filter~\cite{Savitzky1964} to the waveform data. We observe that the tail amplitudes are generally larger for anti-aligned spin configurations. Moreover, the tail amplitudes increase systematically with increasing eccentricity. These trends are consistent across all spin-weighted spherical harmonic modes studied here, thereby generalizing the results previously reported in Refs.~\cite{Islam:2024vro,DeAmicis:2024not}. Furthermore, we also observe that the tail amplitudes are quite similar for modes with the same $\ell$.

We then compute the local power-law exponent $p_{\rm local}(t)$ directly from the mode amplitudes as
\begin{equation}
p_{\rm local}(t) = \frac{d\log |\psi_{4,\ell m}(t)|}{d\log t}.
\end{equation}
Figure~\ref{fig:multi_mode_p_local} shows the estimated $p_{\rm local}(t)$ for all six modes and for all cases considered in this paper. We find that, as time progresses, typically around $t \sim 10^{4}$, $p_{\rm local}(t)$ approaches the expected asymptotic value of $-(\ell + 4)$. This behavior is most evident for $\ell = 2$ and $\ell = 3$. For $\ell = 4$, we observe enhanced numerical noise both in the amplitudes and in the estimated local power-law exponent. Consistent with our findings from the amplitude evolution, the estimated $p_{\rm local}(t)$ values also exhibit qualitatively similar behavior among modes with the same $\ell$. 

\begin{figure}
\includegraphics[width=\columnwidth]{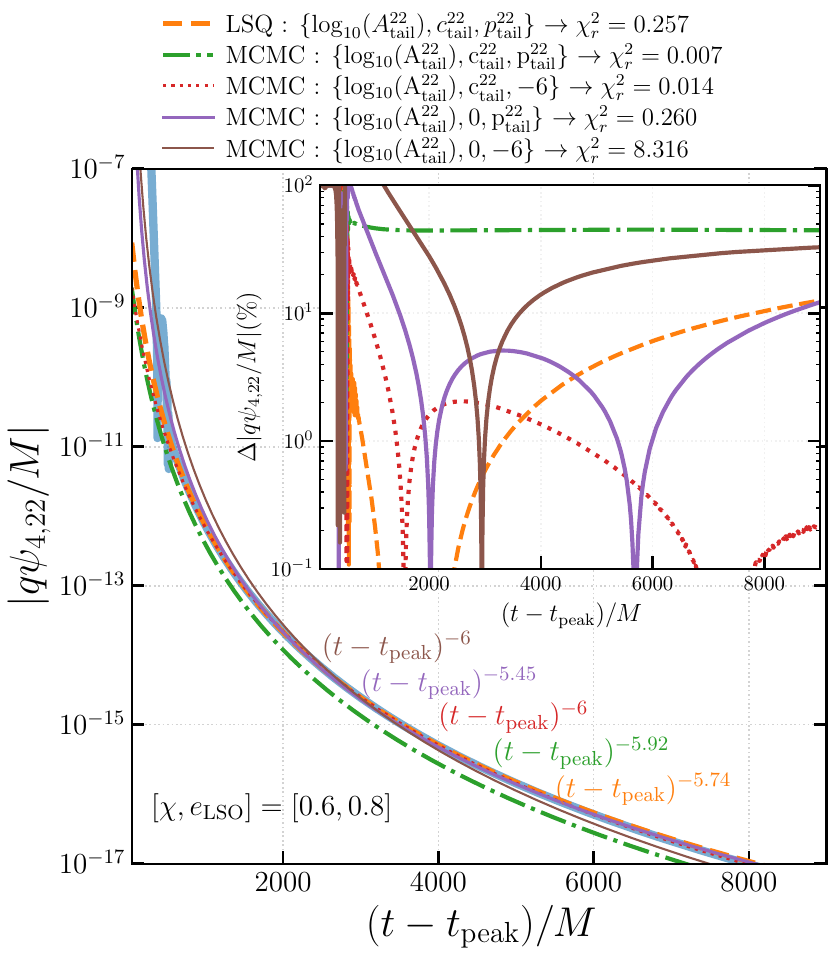}
\caption{We show the $(2,2)$ mode tail amplitude along with the best-fit predictions obtained from different fit assumption and techniques for a binary with $[\chi, e_{\rm LSO}]=[0.6,0.8]$. The analysis is performed over the time window $t = [1200, 8000]M$. Furthermore, we show the residuals for each fit in the insets and report the corresponding reduced $\chi^2$ values in the legends. For reference, we also show the overall best-fit power-law color-coded to match the corresponding best-fit lines. More details are in Section~\ref{sec:example_bbh}.}
\label{fig:example_22_tail_different_fits}
\end{figure}

\subsection{Representative case: $(\chi, e_{\rm LSO}) = (0.6, 0.8)$}
\label{sec:example_bbh}
After inspecting all of our post-merger simulations, we select one representative case with one of the longest evolutions available. This binary is characterized by $(\chi, e_{\rm LSO}) = (0.6, 0.8)$ and tracks the post-merger dynamics up to $9000M$ after the merger. In Figure~\ref{fig:example_hms_tail}, we show the post-merger amplitudes for all six modes considered in this paper. For the $\ell = 2$ and $\ell = 3$ modes, we extract clean and well-behaved tails extending to the end of the simulation. For the $\ell = 4$ modes, we also observe the onset of tail behavior; however, around $t \sim 4200M$, the amplitudes reach the numerical noise floor of the simulation, preventing us from probing the late-time regime further. 

\begin{figure*}
    \centering
    \subfigure[]{
        \includegraphics[width=0.48\textwidth]{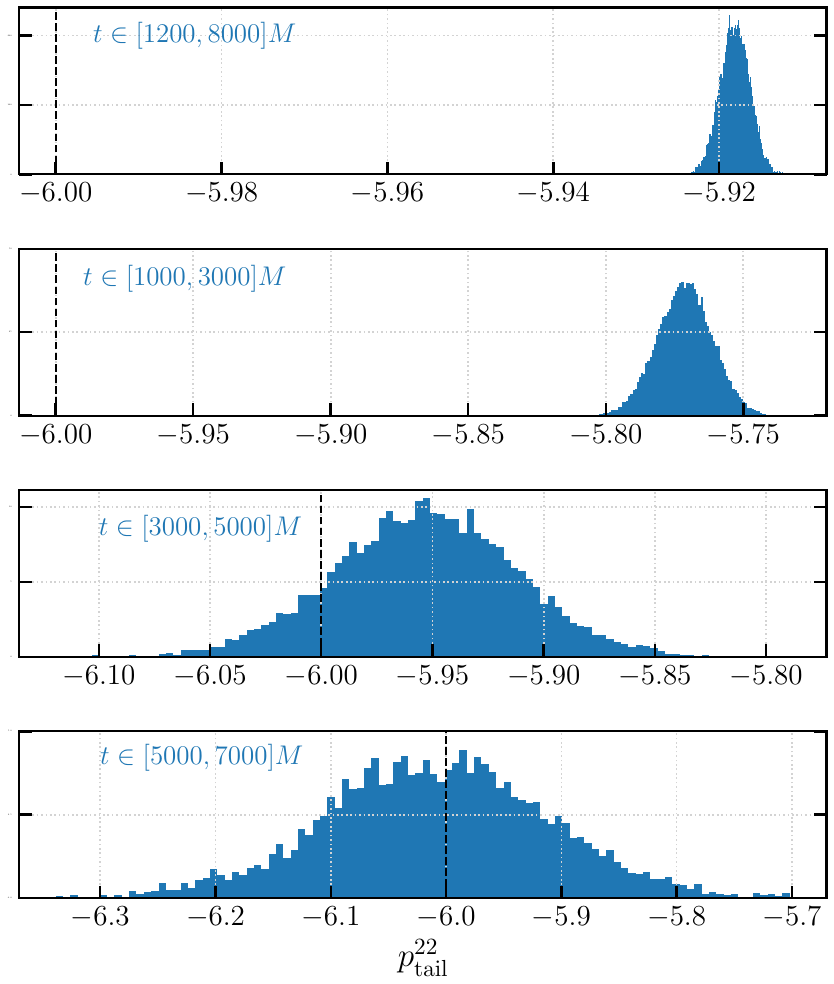}
        \label{fig:example_22_ptail_hists_diff_times}
    }
    \subfigure[]{
        \includegraphics[width=0.48\textwidth]{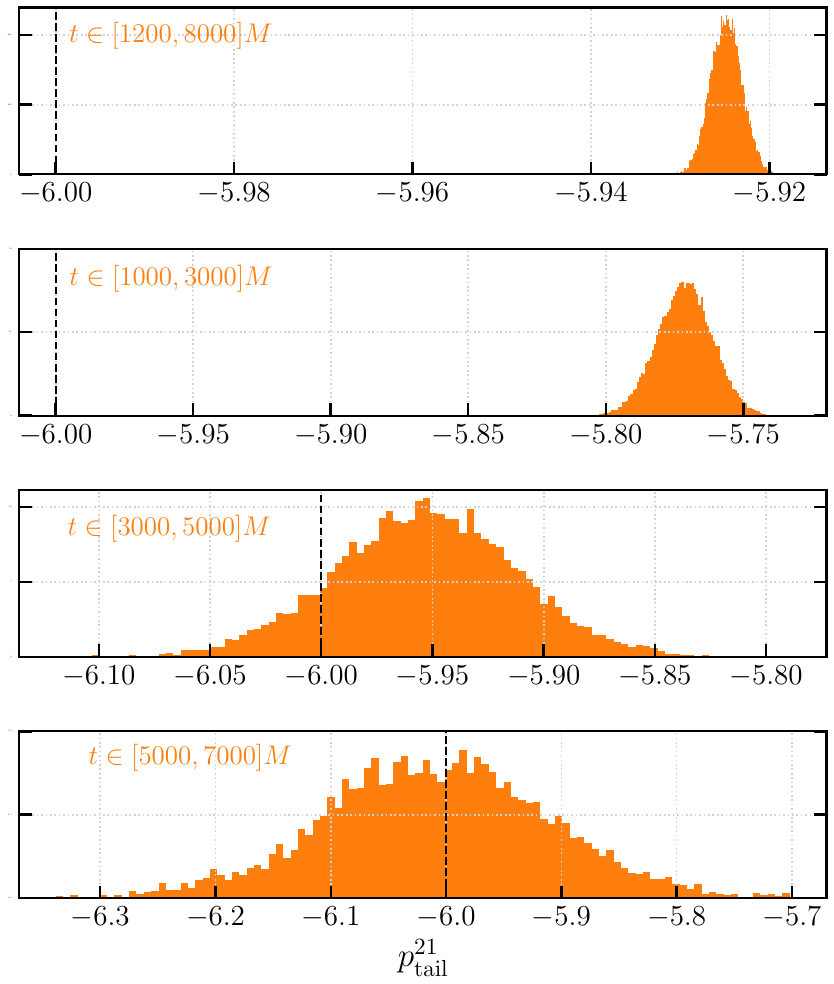}
        \label{fig:example_21_ptail_hists_diff_times}
    }
    \subfigure[]{
        \includegraphics[width=0.48\textwidth]{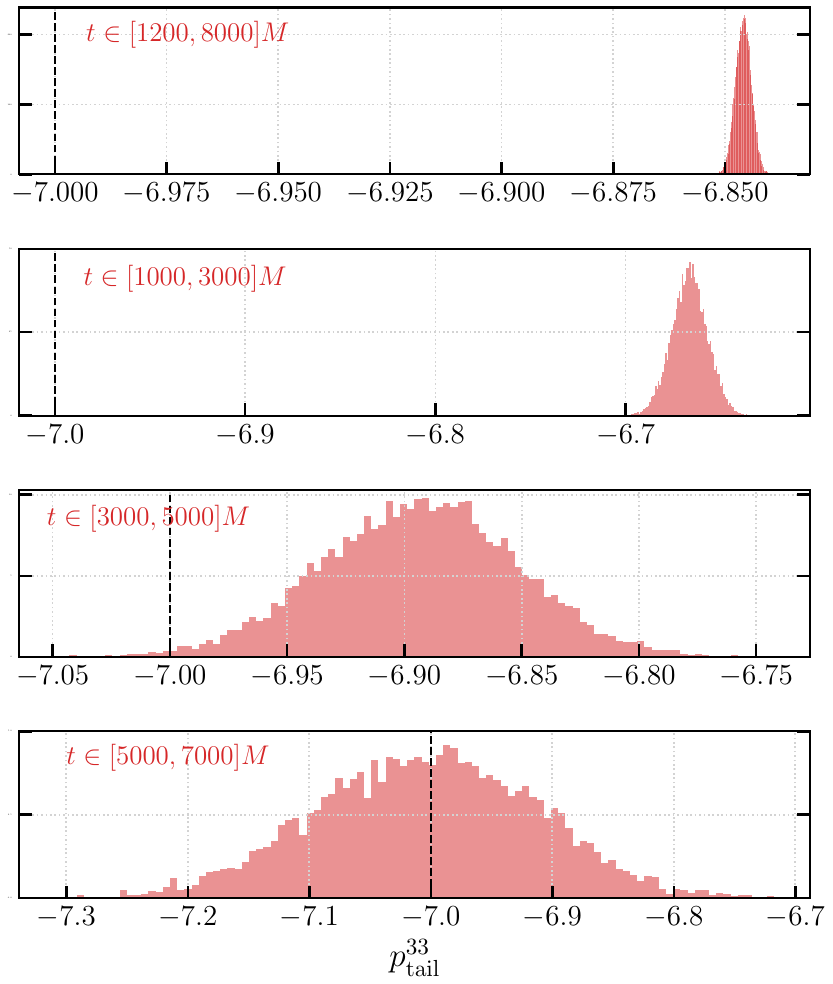}
        \label{fig:example_33_ptail_hists_diff_times}
    }
    \subfigure[]{
        \includegraphics[width=0.48\textwidth]{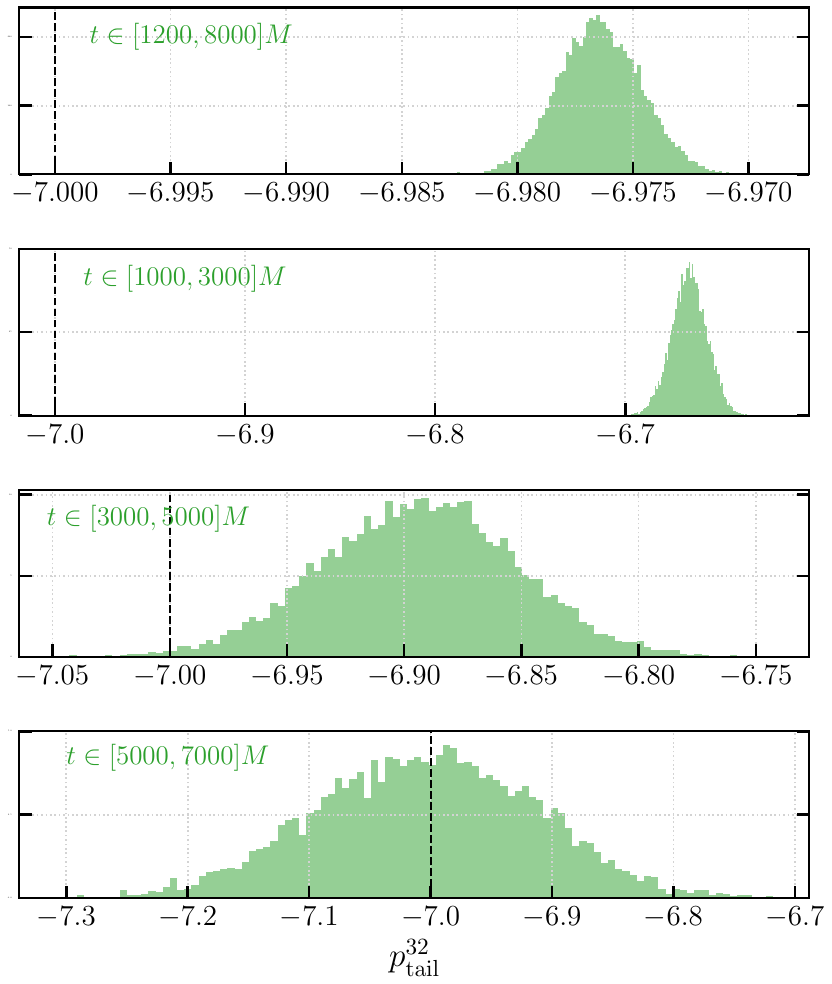}
        \label{fig:example_32_ptail_hists_diff_times}
    }
    \caption{We show the tail exponents $p_{\mathrm{tail}}^{\ell m}$ obtained using different fitting time windows for the $(2,2)$ (upper left), $(2,1)$ (upper right), $(3,3)$ (lower left), and $(4,4)$ (lower right) modes for a binary characterized by $[\chi, e_{\rm LSO}] = [0.6, 0.8]$. For reference, the expected asymptotic values are shown as black dashed lines. As the fitting window shifts to later times, the extracted tail exponents approach their predicted asymptotic values. More details are in Section~\ref{sec:example_bbh}.}
    \label{fig:example_ptail_hists_diff_times}
\end{figure*}

We find that the tail amplitudes are quite similar for modes with the same $\ell$ values, indicating that the late-time decay is primarily governed by the multipolar structure rather than the azimuthal index $m$. Furthermore, we observe that the onset of the tail occurs at different times for different modes. Identifying the precise start time of the tail is generally challenging; however, it is known that the tail phase begins when the instantaneous frequency of a mode, computed as $\omega_{\ell m} = d\phi_{\ell m}/dt$, approaches zero. For our simulation, we find that $\omega_{\ell m}$ reaches zero at different times for different modes. In particular, the tails in the $\ell = 2$ modes start earlier than those in the higher-$\ell$ modes (see the inset of Figure~\ref{fig:example_hms_tail}).

Next, we use \texttt{gwtails}, employing the MCMC sampler \texttt{emcee}, to fit the post-merger amplitudes using our tail model described in Eq.~(\ref{eq:tailmodel}). For the $\ell = 2$ and $\ell = 3$ modes, we perform the fit over the time window $t \in [1200, 8000]M$. We exclude the final $1000M$ of the evolution to provide an additional safeguard against numerical noise contaminating the results. For the $\ell = 4$ modes, we perform the fit over the time window $t \in [1200, 4000]M$. The best-fit tail curves, along with the corresponding power-law exponents, are shown in Figure~\ref{fig:example_hms_tail}. 

In Figure~\ref{fig:example_tail_mcmc}, we present the posterior distributions of the best-fit tail parameters $A_{\mathrm{tail}}^{\ell m}$, $c_{\mathrm{tail}}^{\ell m}$, and $p_{\mathrm{tail}}^{\ell m}$ for four representative modes: $(2,2)$, $(2,1)$, $(3,3)$, and $(4,4)$. The peak values of the posteriors for the power-law index $p_{\mathrm{tail}}^{\ell m}$ are found to be approximately $-5.92$ for the $(2,2)$ mode, $-5.925$ for $(2,1)$, $-6.82$ for $(3,3)$, and $-7.86$ for $(4,4)$. Notably, these values are in close agreement with the expected asymptotic behavior $p_{\mathrm{tail}} = -(\ell + 4)$. 

The parameter $A_{\mathrm{tail}}^{\ell m}$ can be loosely interpreted as being inversely proportional to the tail amplitude, while $c_{\mathrm{tail}}^{\ell m}$ is related to the onset time of the tail regime. We observe that $A_{\mathrm{tail}}^{\ell m}$ systematically increases from $\ell = 2$ to $\ell = 4$, indicating that the actual tail amplitude decreases with increasing $\ell$. Similarly, the time-shift parameter $c_{\mathrm{tail}}^{\ell m}$ decreases with $\ell$, suggesting that higher-$\ell$ modes enter the tail regime later. Furthermore, all best-fit parameters remain consistent between the $(2,2)$ and $(2,1)$ modes (Figure~\ref{fig:example_tail_mcmc}).

Figure~\ref{fig:example_tail_mcmc} also provides valuable insights into the correlations among different fit parameters. These correlations can be understood by examining the two-dimensional posterior distributions between different pairs of tail parameters, namely $A_{\mathrm{tail}}^{\ell m}$, $c_{\mathrm{tail}}^{\ell m}$, and $p_{\mathrm{tail}}^{\ell m}$. We find that the qualitative nature of these correlations remains consistent across all spin-weighted spherical harmonic modes considered. The scatter plots reveal that larger values of $\log_{10}A_{\mathrm{tail}}^{\ell m}$ tend to correlate with smaller values of $p_{\mathrm{tail}}^{\ell m}$, thereby pushing it away from the expected asymptotic value of $-6$. Similarly, larger $\log_{10}A_{\mathrm{tail}}^{\ell m}$ values also favor larger $c_{\mathrm{tail}}^{\ell m}$, implying that the onset of the tail occurs at later times. This trend is unsurprising, as $\log_{10}A_{\mathrm{tail}}^{\ell m}$ is inversely related to the amplitude at the beginning of the tail. Furthermore, we observe a negative correlation between $c_{\mathrm{tail}}^{\ell m}$ and $p_{\mathrm{tail}}^{\ell m}$, indicating that fixing $c_{\mathrm{tail}}^{\ell m} = 0$ prior to fitting will inevitably bias the best-fit $p_{\mathrm{tail}}^{\ell m}$ away from the expected asymptotic behavior of $-(\ell + 4)$.

In Figure~\ref{fig:example_ptail_combined}, we show the evolution of the local tail exponent $p_{\rm local}(t)$ alongside the best-fit posterior distributions of $p_{\mathrm{tail}}^{\ell m}$ obtained from our MCMC fits. For each spin-weighted spherical harmonic mode, $p_{\rm local}(t)$ asymptotically approaches the theoretically expected value of $-(\ell + 4)$ as time progresses.
We note that the best-fit values of $p_{\mathrm{tail}}^{\ell m}$ are significantly closer to the theoretically expected asymptotic values than the late-time estimates obtained directly from $p_{\rm local}(t)$. This indicates that while $p_{\rm local}(t)$ provides a convenient diagnostic for identifying the asymptotic decay behavior, procedures such as a LSQ or MCMC fit are essential for accurately determining the tail exponent. Second, modes with the same $\ell$ exhibit both similar time evolution in their local exponents and consistent posterior distributions for $p_{\mathrm{tail}}^{\ell m}$. For $\ell = 3$, we observe a slight discrepancy between the $p_{\mathrm{tail}}^{\ell m}$ values inferred from the $(3,3)$ and $(3,2)$ modes. We attribute this difference to enhanced numerical noise present in the late-time portion of the waveform, which is also visible in the evolution of $p_{\rm local}(t)$. 

\subsubsection{Effect of fitting ansatz}
Next, we focus on the dominant $(2,2)$ mode and investigate how the inferred tail parameters depend on different fitting choices. We repeat the tail fits under several conditions: first by fixing the power-law exponent to its asymtotically expected value $p_{\mathrm{tail}}^{22} = -6$, then by fixing the time-shift parameter $c_{\mathrm{tail}}^{22} = 0$, and finally by fixing both simultaneously. As a baseline, we compare these results against fits performed with all three parameters $\{A_{\mathrm{tail}}^{22}, c_{\mathrm{tail}}^{22}, p_{\mathrm{tail}}^{22}\}$ allowed to vary freely, using both a LSQ approach and an MCMC sampler. For each fit, we compute the reduced chi-squared statistic $\chi^2_{\mathrm{red}}$ to assess the goodness of fit, and we also evaluate the differences between the actual tail amplitudes and those predicted by the best-fit models.

We present the results in Figure~\ref{fig:example_22_tail_different_fits}. We find that the inferred tail exponent $p_{\mathrm{tail}}^{22}$ varies significantly across different fitting strategies. In particular, we obtain values closest to the theoretically expected asymptotic value of $-6$ when all three parameters are allowed to vary simultaneously. Moreover, the MCMC-based fits yield exponents systematically closer to the expected value than those obtained from the LSQ approach. Consistent with this, the smallest $\chi^2_{\mathrm{red}}$ is achieved for the MCMC fit with all three tail parameters free, yielding $\chi^2_{\mathrm{red}} = 0.007$. We find that the fit residuals are mostly below $1\%$ for this case. In contrast, the poorest fit is obtained when both $c_{\mathrm{tail}}^{22} = 0$ and $p_{\mathrm{tail}}^{22} = -6$ are fixed, in which case the reduced chi-squared rises to $\chi^2_{\mathrm{red}} = 8.316$. The fit residuals are found to be as large as $30\%$ for this case.

\subsubsection{Effect of fitting time window}
We next investigate whether the late-time tail exponent is already close to its theoretically expected asymptotic value and whether fits performed over a wide time window might be biased by early-time contributions. To probe this possibility, we repeat the tail fits for the $(2,2)$, $(2,1)$, $(3,2)$, and $(3,3)$ modes using three progressively later time windows: $t = [1000, 3000]M$, $t = [3000, 5000]M$, and $t = [5000, 7000]M$. The resulting best-fit values of $p_{\mathrm{tail}}^{\ell m}$ are presented in Figure~\ref{fig:example_ptail_hists_diff_times}.
Our results show that the tail exponents inferred from the broad time window $t = [1200, 8000]M$ are indeed contaminated by early-time data, leading to slight deviations from the true asymptotic values. As the fitting window is shifted towards later times, the recovered $p_{\mathrm{tail}}^{\ell m}$ values consistently move closer to their expected asymptotic limits. In particular, fits restricted to the latest time window $t = [5000, 7000]M$ yield $p_{\mathrm{tail}}^{22} = -6$, $p_{\mathrm{tail}}^{21} = -6$, $p_{\mathrm{tail}}^{32} = -6$, and $p_{\mathrm{tail}}^{33} = -6$, thereby recovering the theoretically predicted values to within $\sim 3\%$ accuracy. This demonstrates that early-time data can bias the estimation of the tail exponent and highlights the importance of focusing on the late-time regime to accurately capture the asymptotic power-law behavior.

\subsection{Comparison with other works}
\label{sec:comparison}
\begin{figure}
\includegraphics[width=\columnwidth]{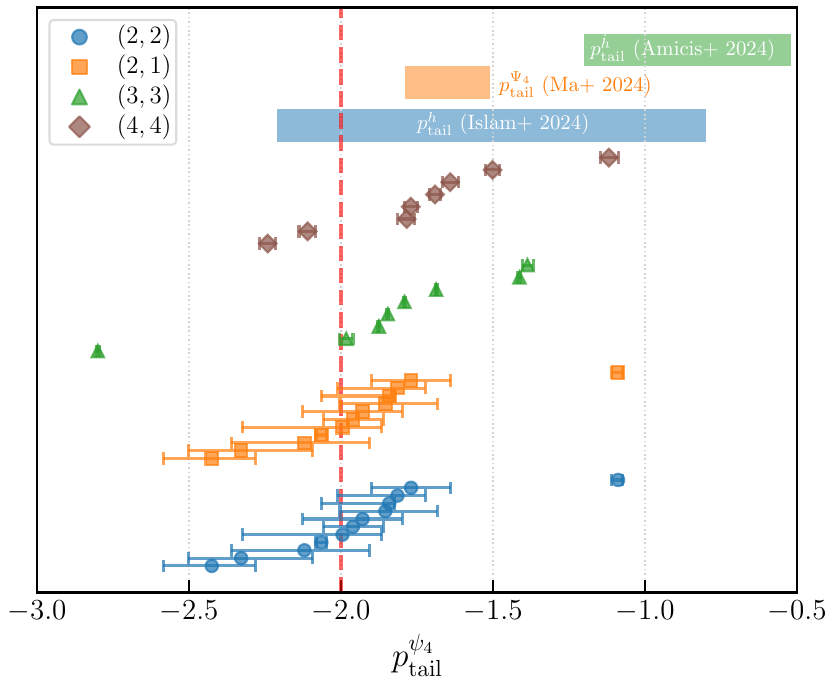}
\caption{We show the median common tail exponent $p_{\mathrm{tail}}^{\Psi_4}$, along with the corresponding $90\%$ credible interval, computed from different spin-weighted spherical harmonic modes for all binaries considered in Section~\ref{sec:all_bbhs}. For reference, the vertical dashed red line indicates the common expected asymptotic value of $-2$. For comparison, we also show the range of values reported in recent literature, that employ either NR or perturbative data, as bands. More details are in Section~\ref{sec:comparison}.}
\label{fig:common_ptail}
\end{figure}

Finally, to summarize our results in a more compact manner, we define a \emph{common tail exponent} for all spherical harmonic modes of $\Psi_4$, denoted by $p_{\rm tail}^{\Psi_4}$, as
\begin{equation}
p_{\rm tail}^{\Psi_4} = p^{\ell m}_{\rm tail} + \ell + 2\,.
\end{equation}
We include the $\Psi_4$ term in the superscript to facilitate comparison with the other estimates discussed below.
With this definition, the expected asymptotic value of the common tail exponent is $-2$ for all modes of $\Psi_4$. In Figure~\ref{fig:common_ptail}, we show the median value of $p_{\mathrm{tail}}^{\Psi_4}$, along with the corresponding $90\%$ credible intervals, computed from four spin-weighted spherical harmonic modes, $(2,2)$, $(2,1)$, $(3,3)$, and $(4,4)$, for all binaries considered in this work. For visual clarity, we arrange the results from the smallest to the largest median value. We find that, except for a few cases where enhanced numerical noise may be present, all binaries consistently yield a common tail exponent close to the expected value across different modes.

To facilitate a direct comparison with the tail exponents reported in recent studies~\cite{Ma:2024hzq,DeAmicis:2024not,DeAmicis:2024eoy,Islam:2024vro}, which compute these quantities from the spin-weighted spherical-harmonic modes $h_{\ell m}$ of the GW strain $h$ (defined in Eq.~(\ref{eq:strain})), we define an alternative \emph{common tail exponent} as
\begin{equation}
p_{\rm tail}^{h} = p_{\rm tail}^{h,\ell m} + \ell\,.
\end{equation}
With this definition, both $p_{\rm tail}^{\Psi_4}$ and $p_{\rm tail}^{h}$ share the same expected asymptotic value of $-2$. Likewise, when extracting tails from the news amplitude $|\dot{h}|$, one can define
\begin{equation}
p_{\rm tail}^{\dot{h}} = p_{\rm tail}^{\dot{h},\ell m} + \ell + 1\,,
\end{equation}
which allows for a unified comparison of tail exponents derived from different waveform quantities reported in the literature. 
In Figure~\ref{fig:common_ptail}, we present the median common tail exponent $p_{\mathrm{tail}}^{\Psi_4}$, along with the corresponding $90\%$ credible intervals, computed from different spin-weighted spherical harmonic modes for all binaries considered in this work. For comparison, we also display the ranges of values reported in recent literature that employ either NR simulations or perturbative data and use either the strain or the news amplitude to compute tail exponents. 

Our common tail exponents $p_{\rm tail}^{\Psi_4}$ lie significantly closer to the expected asymptotic value than those reported in other analyses. For example, Ref.~\cite{Ma:2024hzq}, which studied head-on and quasi–head-on BBH collisions using NR simulations, reports $p_{\rm tail}^{\Psi_4}$ values in the range $[-1.75, -1.51]$. Similarly, Ref.~\cite{DeAmicis:2024not}, that utilizes NR data too, finds $p_{\rm tail}^{\dot{h}}$ values between $-1.2$ and $-0.52$, substantially smaller than the theoretical expectation of $-2$. These deviations are likely due to the fact that both of these NR simulations could probe only the early part of the tail, typically extending to $\sim 400M$ after merger. By contrast, our previous perturbative analysis~\cite{Islam:2024vro}, which extended to $\sim 1000M$ after merger but included only the quadrupolar mode, yielded $p_{\rm tail}^{\Psi_4}$ values in the range $[-2.1, -0.75]$, showing partial overlap with the results presented here.

\section{Intermediate tails}
\label{sec:intermediate_tails}
\begin{figure}
\includegraphics[width=\columnwidth]{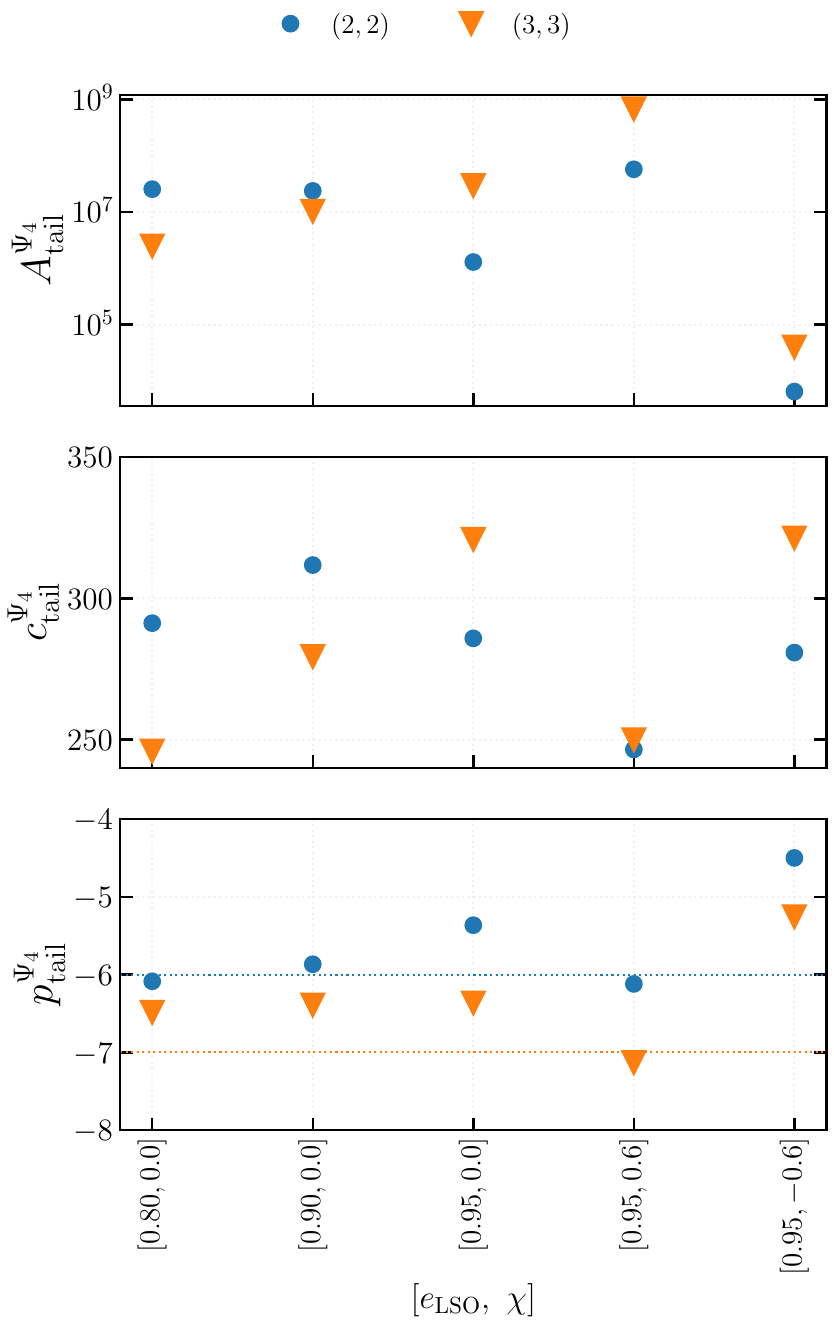}
\caption{We show the best-fit tail parameters for two representative modes, $(2,2)$ and $(3,3)$, at intermediate times (following the oscillatory phase immediately after the QNM-dominated regime) for binaries with varying spin and eccentricity. Dotted horizontal lines indicate the expected asymptotic values of $p_{\rm tail}^{\Psi_4}=-6$ and $p_{\rm tail}^{\Psi_4}=-7$ for the $(2,2)$ and $(3,3)$ modes, respectively. More details are in Section~\ref{sec:intermediate_tails}.}
\label{fig:intermediate_fits}
\end{figure}

While most of this paper focuses on late-time tails, Ref.~\cite{Islam:2024vro} explored the behavior of tails up to $800M$--$1000M$ after merger, with particular emphasis on the intermediate regime, and also presented fits for the tail amplitude, power-law exponent, and time shift of the $(2,2)$ mode as functions of eccentricity and spin. In this work, for the sake of completeness, we fit the tails (using a least-squares approach) for the $(2,2)$ and $(3,3)$ modes just after the oscillatory regime, allowing for a buffer of $\sim 100M$, and present our results in Fig.~\ref{fig:intermediate_fits}. These fits are obtained using $t\in[600,1000]M$. Consistent with Ref.~\cite{Islam:2024vro}, we find that the exponent $p_{\rm tail}^{\Psi_4}$ increases with eccentricity at fixed spin. At fixed eccentricity, the exponent is larger for large negative spin compared to the nonspinning case, and smaller for large positive spin, for both modes. Furthermore, as expected, the tail exponent is more negative for the $(3,3)$ mode than for the $(2,2)$ mode. Hints of a similar trend are also observed for the time-shift parameter $c_{\rm tail}^{\Psi_4}$. For most cases, $c_{\rm tail}^{\Psi_4}$ lies in the range $200M$--$350M$, consistent with results obtained from Bayesian analyses using late-time tail data, as well as those reported in Ref.~\cite{Islam:2024vro}. Similar behavior is observed for the other modes.

\section{Discussion \& Conclusion}
In this paper, we present a detailed analysis of late-time GW tails for 15 spin-aligned eccentric BBH mergers with varying spins $\chi = [-0.9, -0.6, 0.0, 0.6, 0.9]$ and eccentricities at LSO $e_{\rm LSO} = [0.8, 0.9, 0.95]$ at an intermediate mass ratio of $q = 1000$, using high-accuracy ppBHPT simulations. Within ppBHPT, we model these BBHs as a test-particle orbiting on the equatorial plane of a Kerr BH.
Our ppBHPT simulations mark a significant advancement over those presented in Ref.~\cite{Islam:2024vro}, incorporating several numerical improvements that now enable us to probe post-merger radiation not only in the dominant $(2,2)$ mode but also in several higher-order modes, including $(2,1)$, $(3,2)$, $(3,3)$, $(4,3)$, and $(4,4)$. These simulations provide a tool to investigate late-time tail behavior. Furthermore, to ensure a more robust characterization of the tail dynamics, we employ a Bayesian parameter estimation approach using the \texttt{emcee} MCMC sampler.

Our analysis reveals several key results. First, we find that the late-time power-law exponents of the waveform decay approach the predicted asymptotic values, $p = -\ell - 4$, for $\Psi_{4,\ell m}$. When the analysis is restricted to the latest portion of the signal, we recover these theoretical values exactly, illustrating that early-time data can bias the estimation of tail exponents. Second, we confirm numerically that modes with the same spin-weighted spherical harmonic index $\ell$ share identical late-time tail exponents, indicating that variations in $m$ do not affect the asymptotic tail behavior.
The full analysis framework developed in this work is publicly available through the \texttt{gwtails} Python package ({\href{https://github.com/tousifislam/gwtails}{https://github.com/tousifislam/gwtails}).

In the future, we aim to extend this study to precessing eccentric binaries to investigate whether couplings between eccentricity and precession further enhance the tail amplitudes. It would also be valuable to push current NR simulations to significantly later times (beyond $500M$)~\cite{DeAmicis:2024eoy,Ma:2024hzq} than is currently possible. Such longer evolutions would help validate several of the results presented here, particularly by enabling a more detailed exploration of the tails in higher-order modes using NR data. Apart from these, it will be helpful to extend the analysis presented in Ref.~\cite{Becker:2025inprep} to higher order modes and study how late-time dynamics as controlled by the anomaly and eccentricity affects the tails in a more comprehensive manner.
Another interesting direction is to develop a template for the best-fit tail parameters as a function of spin and eccentricity, which would greatly facilitate incorporating late-time tail behavior into ringdown modeling. We leave these investigations for future.

\begin{acknowledgments}
We thank Scott Field for helpful discussions, and Devin R. Becker and Scott A. Hughes for their comments on the manuscript.
T.I. is supported in part by the National Science Foundation under Grant No. NSF PHY-2309135 and the Gordon and Betty Moore Foundation Grant No. GBMF7392. G.K. acknowledges support from NSF Grants No. PHY-2307236 and DMS-2309609.

Use was made of computational facilities purchased with funds from the National Science Foundation (CNS-1725797) and administered by the Center for Scientific Computing (CSC). The CSC is supported by the California NanoSystems Institute and the Materials Research Science and Engineering Center (MRSEC; NSF DMR 2308708) at UC Santa Barbara. 
This work was partly supported by UMass Dartmouth's Marine and Undersea Technology (MUST) research program funded by the Office of Naval Research (ONR) under grant no. N00014-23-1-2141.
Some computations were performed on the UMass-URI UNITY HPC/AI cluster at the Massachusetts Green High-Performance Computing Center (MGHPCC).
\end{acknowledgments}

\bibliography{References}

\appendix

\section{All binaries}
\label{sec:all_bbhs}
We now present a comprehensive analysis of the tail parameters extracted from all binary simulations considered in this work, across the different spin-weighted spherical harmonic modes. First, we confirm without explicitly showing it here that the findings discussed in the previous subsection are not specific to the representative binary chosen earlier, but rather hold generally across the entire set of simulations. Now, for the sake of completeness, we summarize results for four key modes: $(2,2)$, $(2,1)$, $(3,3)$, and $(4,4)$.
In this section, we adopt a consistent fitting procedure by using a default time window of $t = [1200, 8000]M$ for the post-merger analysis. For simulations that do not extend up to $8000M$ after the merger, we instead use the end time of the available data as the upper bound of the fitting window. This uniform approach enables a direct and meaningful comparison of tail parameters across different binaries and modes.

We present the estimated tail parameters for a total of 12 binary configurations with varying spins and eccentricities: the tail exponent $p_{\mathrm{tail}}^{22}$ in Figure~\ref{fig:22_tails_ptail_all_bbhs}, the time-shift parameter $c_{\mathrm{tail}}^{22}$ in Figure~\ref{fig:22_tails_ctail_all_bbhs}, and the logarithm of the tail amplitude $A_{\mathrm{tail}}^{22}$ in Figure~\ref{fig:22_tails_logAtail_all_bbhs}. 
Furthermore, we repeat the fits for the $(2,2)$ mode tails using only the last $2000M$ of the post-merger data. This allows us to quantify the systematic differences in the inferred tail parameters, particularly $p_{\mathrm{tail}}^{22}$, when using a longer fitting window versus restricting the analysis to the latest times. Consistent with the findings of the previous section, when we limit the fit to the last $2000M$, the estimated $p_{\mathrm{tail}}^{22}$ shifts closer to the expected asymptotic value of $-6$ for all binaries considered. More importantly, in many cases, it becomes fully consistent with $-6$ within the statistical uncertainties.

We find that the extracted tail parameters are broadly consistent across different binaries, showing only minor variations. All of these binaries yield $p_{\mathrm{tail}}^{22}$ close to $-6$.
These small differences can be attributed to two main factors. First, different simulations evolve to different times after merger, leading to slight variations in the fitting windows used for parameter estimation. Second, the level of numerical noise differs across simulations, particularly in the late-time regime, which can affect the precision of the extracted tail parameters.

Similarly, in Figure~\ref{fig:tails_ptail_all_bbhs}, we present the best-fit tail exponents for the $(2,1)$, $(3,3)$, and $(4,4)$ modes across nine binary configurations with varying spins and eccentricities. Three additional binaries were excluded from this analysis due to enhanced numerical noise in the late-time data. We find that the extracted tail exponents cluster closely around the expected asymptotic values: approximately $-6$ for the $(2,1)$ mode, $-7$ for the $(3,3)$ mode, and $-8$ for the $(4,4)$ mode. Consistent results are also obtained for the $(3,2)$ and $(4,3)$ modes, with best-fit exponents of $-7$ and $-8$, respectively (not shown here). Overall, these findings reinforce the conclusions drawn earlier from the representative case, demonstrating that the late-time tail behavior is robust across a range of binary parameters.

\begin{figure*}
    \centering
    \subfigure[]{
        \includegraphics[width=0.98\textwidth]{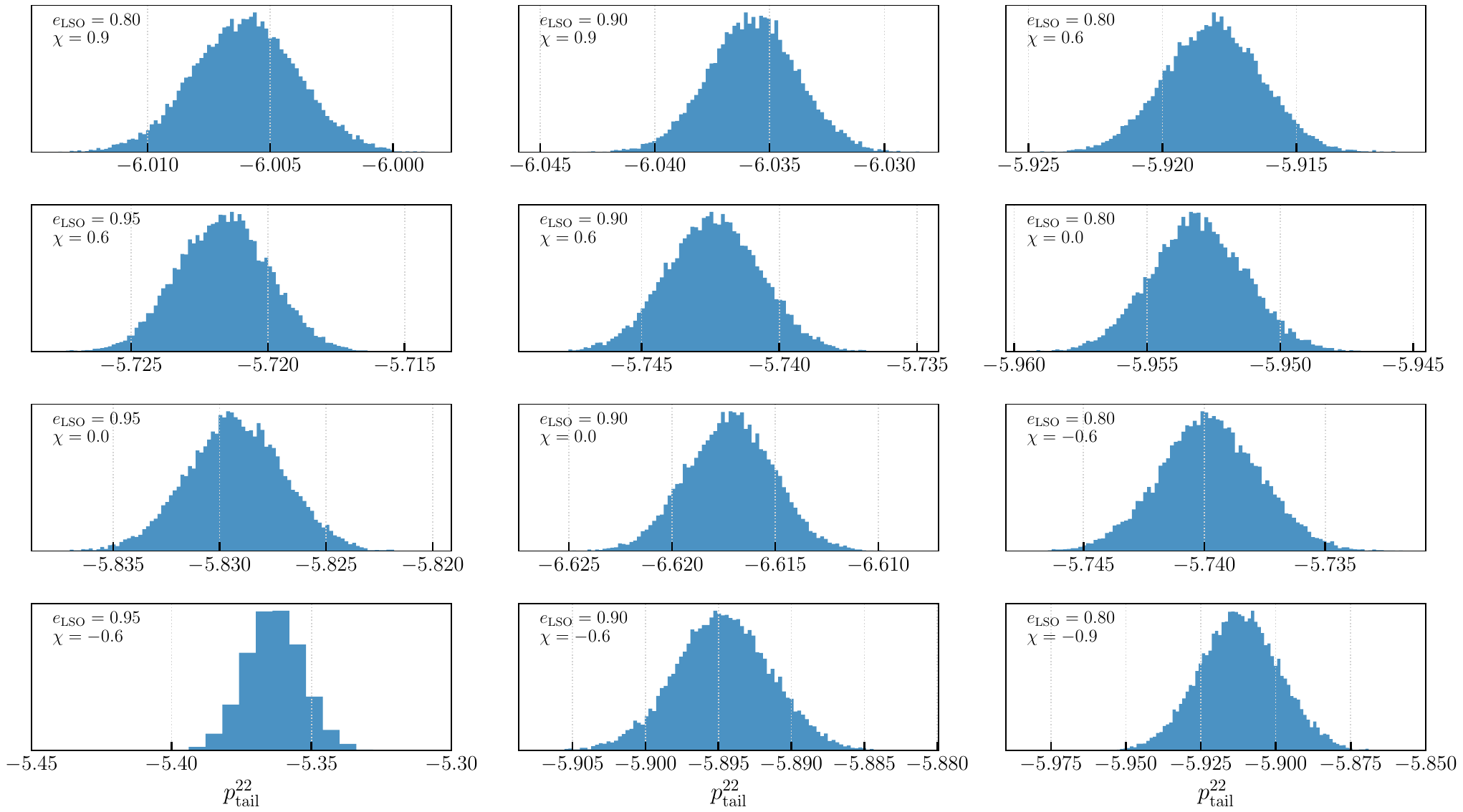}
        \label{fig:22_tails_ptail_all_bbhs}
    }
    \vspace{0.3cm}
    \subfigure[]{
        \includegraphics[width=0.96\textwidth]{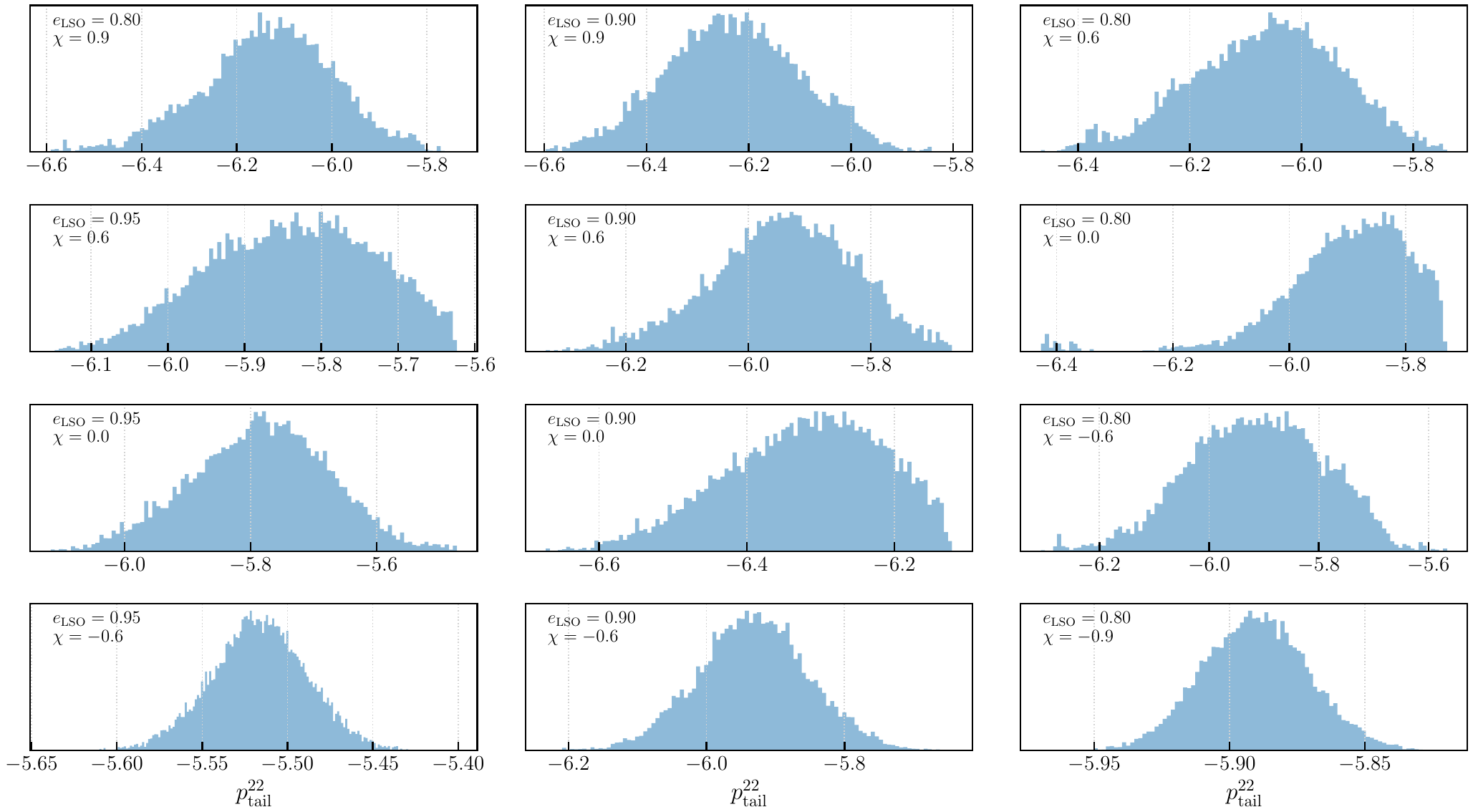}
        \label{fig:22_tails_ptail_all_bbhs_last2000M}
    }
    \caption{We show the best-fit tail exponent $p_{\mathrm{tail}}^{\ell m}$  for the $(2,2)$ mode for binaries with varying spin and eccentricity while using (a) full time window $t=[1200,8000]M$ and (b) last $2000M$ of the post-merger data. We find that the $p_{\mathrm{tail}}^{\ell m}$ values extracted from the later times of the evolution (the last $2000M$ of the evolution) closely match the expected asymptotic value of $-6$. More details are in Section~\ref{sec:all_bbhs}. 
    }
\end{figure*}

\begin{figure*}
    \centering
    \subfigure[]{
        \includegraphics[width=0.98\textwidth]{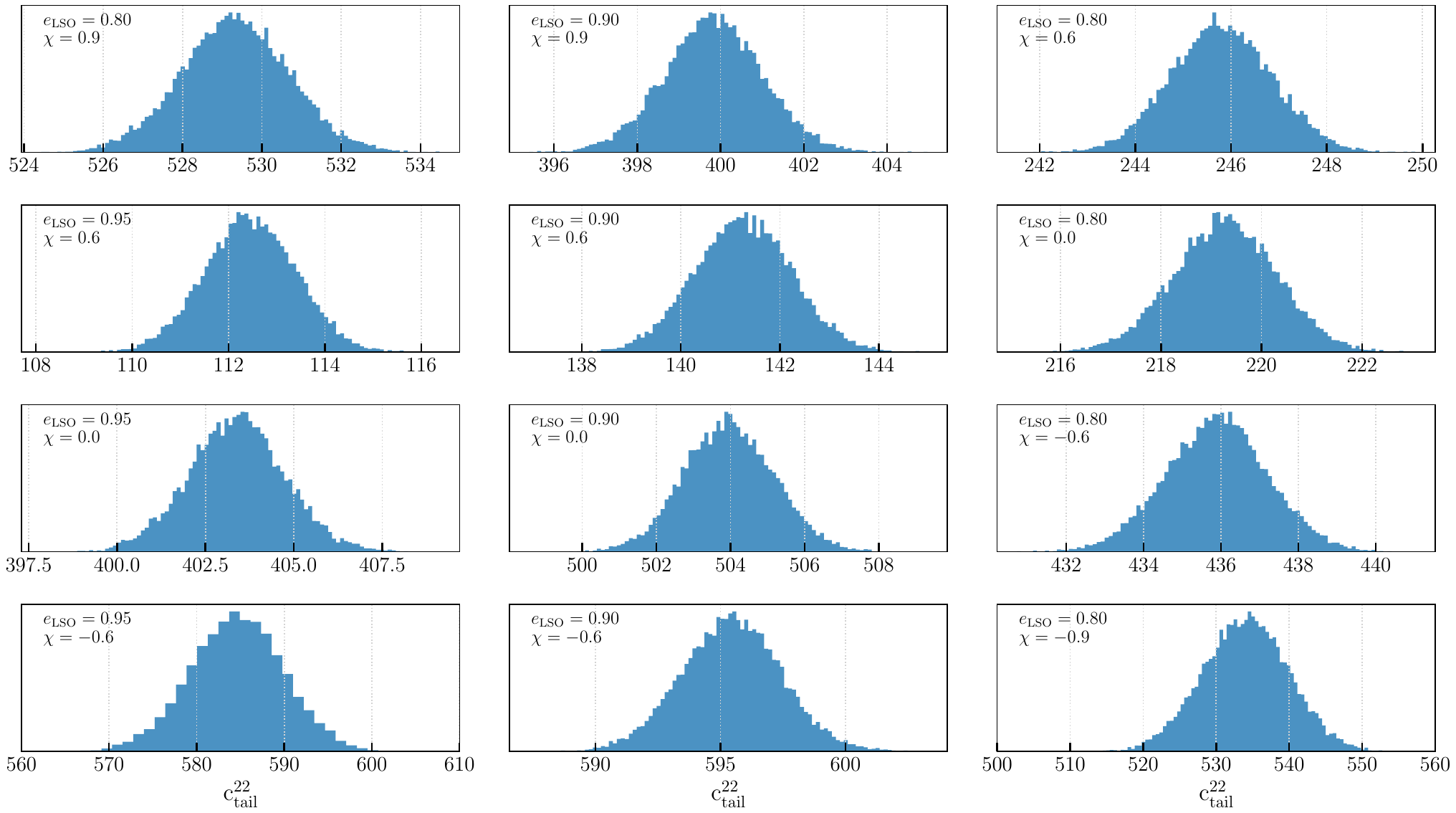}
        \label{fig:22_tails_ctail_all_bbhs}
    }
    \vspace{0.3cm}
    \subfigure[]{
        \includegraphics[width=0.98\textwidth]{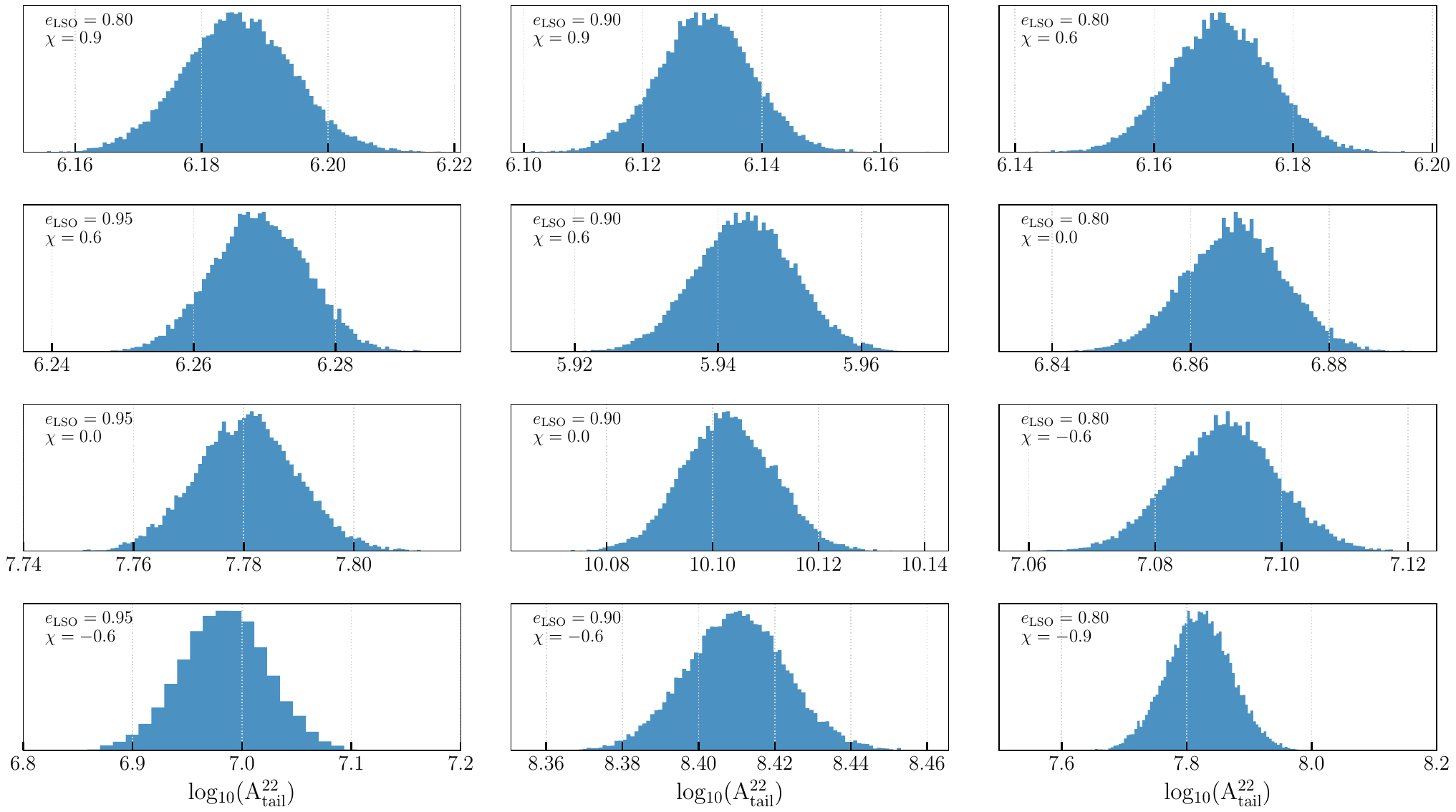}
        \label{fig:22_tails_logAtail_all_bbhs}
    }
    \caption{We show test-fit tail time shifts $c_{\mathrm{tail}}^{\ell m}$ (upper panel) and tail amplitudes $A_{\mathrm{tail}}^{\ell m}$ (lower panel; shown on a logarithmic scale) for the $(2,2)$ mode across binaries with varying spin and eccentricity. The fits are performed using the default time window $t = [1200, 8000]M$. More details are in Section~\ref{sec:all_bbhs}.}
\end{figure*}

\begin{figure*}
    \centering
    \subfigure[]{
        \includegraphics[width=0.93\textwidth]{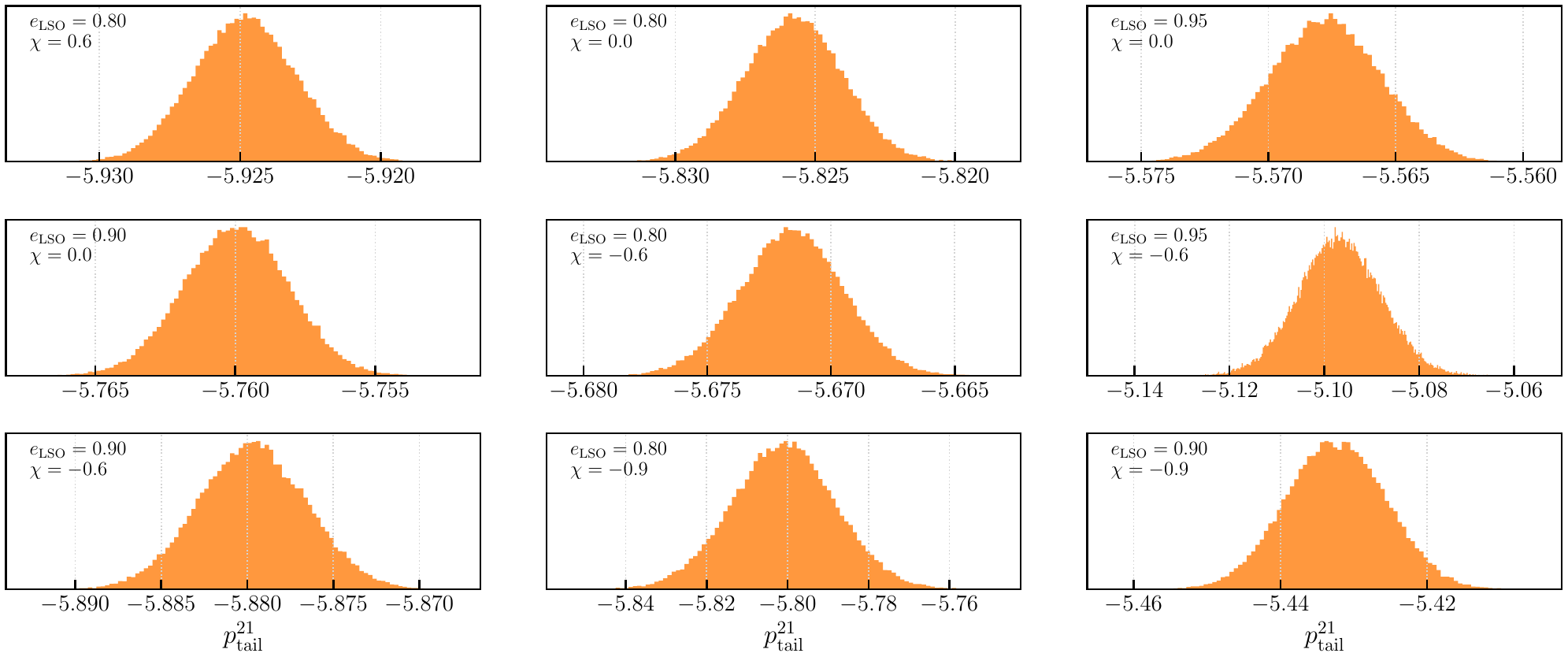}
        \label{fig:21_tails_ptail_all_bbhs}
    }
    \vspace{0.3cm}
    \subfigure[]{
        \includegraphics[width=0.93\textwidth]{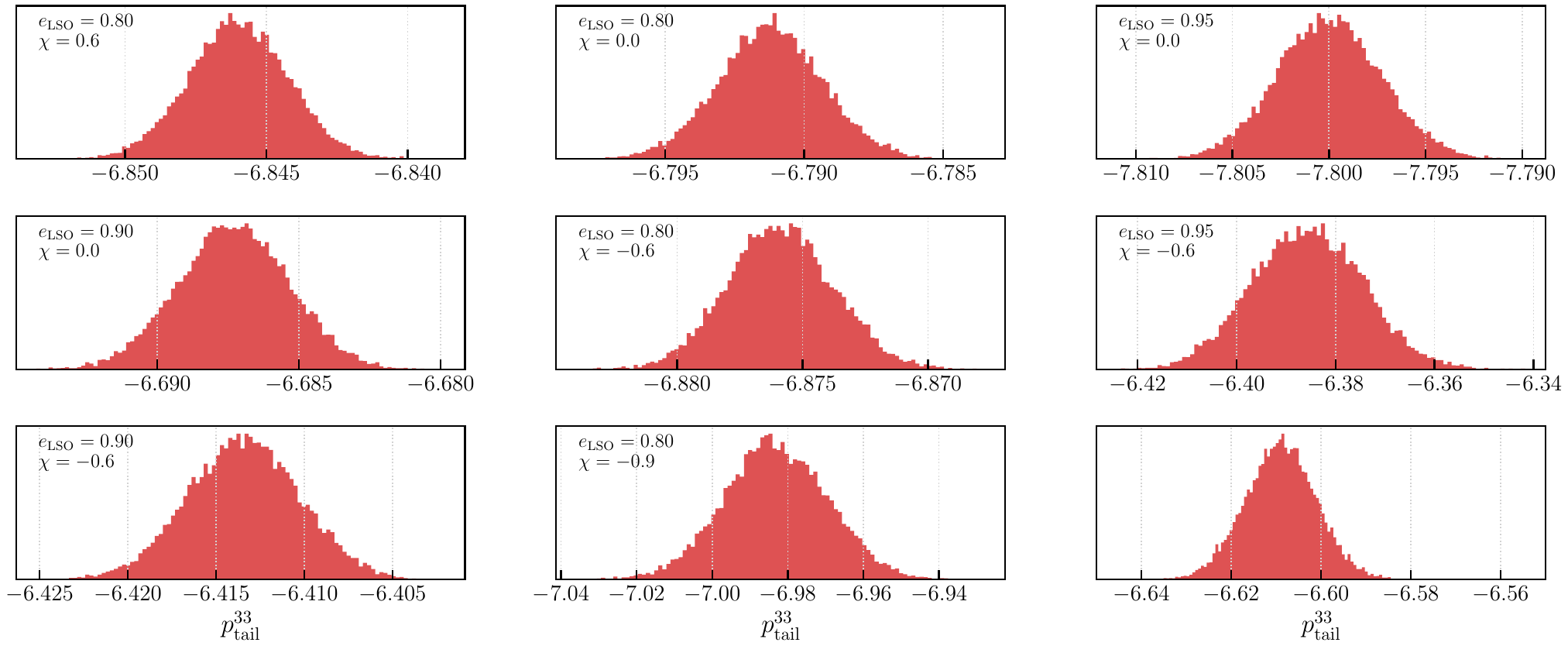}
        \label{fig:33_tails_ptail_all_bbhs}
    }
    \vspace{0.3cm}
    \subfigure[]{
        \includegraphics[width=0.93\textwidth]{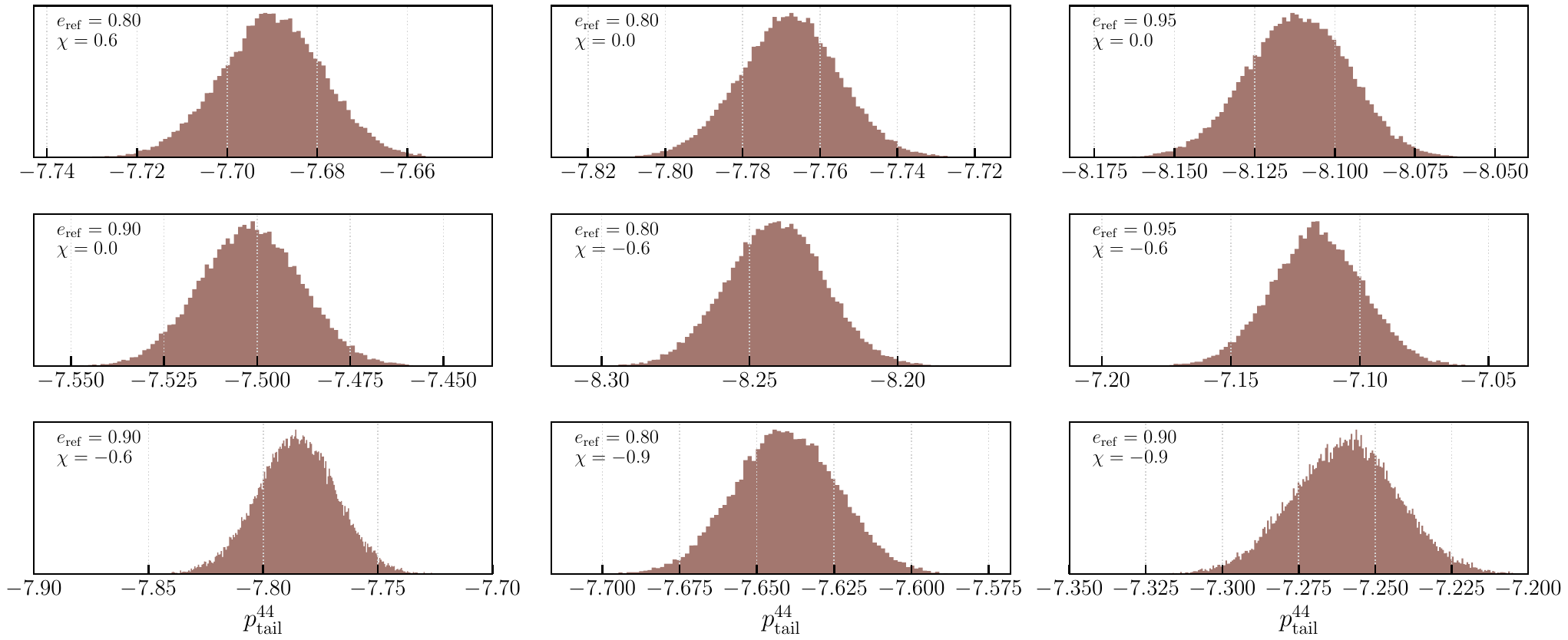}
        \label{fig:44_tails_ptail_all_bbhs}
    }
    \caption{We show test-fit tail exponent $p_{\mathrm{tail}}^{\ell m}$ for the $(2,1)$ (upper panel), $(3,3)$ (middle panel) and $(4,4)$ (lower panel) modes for binaries with varying spin and eccentricity. The fits are performed using the default time window $t = [1200, 8000]M$. More details are in Section~\ref{sec:all_bbhs}.}
    \label{fig:tails_ptail_all_bbhs}
\end{figure*}

\end{document}